\documentclass[acmsmall]{acmart}
\AtBeginDocument{%
  }

\setcopyright{cc}
\setcctype{by}
\acmDOI{10.1145/3798258}
\acmYear{2026}
\acmJournal{PACMPL}
\acmVolume{10}
\acmNumber{OOPSLA1}
\acmArticle{150}
\acmMonth{4}
\received{2025-10-10}
\received[accepted]{2026-02-17}

\usepackage{xspace}
\usepackage[capitalise]{cleveref}
\newcommand{\tool}{\textsc{TransFuzz}\xspace}

\newcommand{\F}{Fig.}

\newcommand{\T}{Table}
\renewcommand{\S}{Sec.}
\newcommand{\A}{Alg.}

\newcommand{\ignore}[1]{}
\newcommand{\mysubref}[2]{\hyperref[#1]{\ref*{#1}(#2)}}

\usepackage{array, multirow, booktabs}
\usepackage{tikz}
\usepackage{xcolor}
\usepackage{pifont}
\usepackage{threeparttable}
\usepackage{xspace}
\usepackage{makecell}

\newcolumntype{L}[1]{>{\raggedright\arraybackslash}m{#1}}
\newcolumntype{C}[1]{>{\centering\arraybackslash}m{#1}}

\usepackage{colortbl}
\usepackage{xcolor}
\usepackage{algpseudocode}
\usepackage[ruled,vlined,linesnumbered]{algorithm2e}
\usepackage{graphicx}
\usepackage{wrapfig}
\usepackage{subcaption}

\definecolor{crash}{RGB}{255,204,204}        
\definecolor{cpugpu}{RGB}{204,255,204}       
\definecolor{eagercompile}{RGB}{204,229,255} 
\definecolor{wrongval}{RGB}{255,255,204}     
\definecolor{funfail}{RGB}{204,229,204}      
\definecolor{abnormal}{RGB}{224,204,255}     

\definecolor{mygreen}{RGB}{155,188,167}
\definecolor{myred}{RGB}{228,181,176}

\usepackage[most]{tcolorbox}

\begin{document}

\title{LLM-Powered Silent Bug Fuzzing in Deep Learning Libraries via Versatile and Controlled Bug Transfer}

\author{Kunpeng Zhang}
\orcid{0009-0009-5413-733X}
\affiliation{%
  \institution{Hong Kong University of Science and Technology}
  \city{Hong Kong}
  \country{China}
}
\email{kzhangcl@connect.ust.hk}

\author{Dongwei Xiao}
\authornote{Corresponding author.}
\orcid{0000-0002-4680-5715}
\affiliation{%
  \institution{Hong Kong University of Science and Technology}
  \city{Hong Kong}
  \country{China}
}
\email{dxiaoad@cse.ust.hk}

\author{Daoyuan Wu}
\authornote{Work done while the author was associated with HKUST.}
\orcid{0000-0002-3752-0718}
\affiliation{%
  \institution{Lingnan University}
  \city{Hong Kong}
  \country{China}
}
\email{daoyuanwu@ln.edu.hk}

\author{Shuai Wang}
\orcid{0000-0002-0866-0308}
\affiliation{%
  \institution{Hong Kong University of Science and Technology}
  \city{Hong Kong}
  \country{China}
}
\email{shuaiw@cse.ust.hk}

\author{Jiali Zhao}
\orcid{0009-0007-4389-2925}
\affiliation{%
  \institution{Huawei}
  \city{Shenzhen}
  \country{China}
}
\email{zhaojiali3@huawei.com}

\author{Yuanyi Lin}
\orcid{0000-0003-4069-2550}
\affiliation{%
  \institution{Huawei}
  \city{Shenzhen}
  \country{China}
}
\email{linyuanyi2@huawei.com}

\author{Tongtong Xu}
\orcid{0000-0002-4323-497X}
\affiliation{%
  \institution{Huawei}
  \city{Shenzhen}
  \country{China}
}
\email{xutongtong9@huawei.com}

\author{Shaohua Wang}
\orcid{0000-0001-5777-7759}
\affiliation{%
  \institution{Central University of Finance and Economics}
  \city{Shenzhen}
  \country{China}
}
\email{davidshwang@ieee.org}

\renewcommand{\shortauthors}{K. Zhang, D. Xiao, D. Wu, J. Zhao, Y. Lin, T. Xu, S. Wang, and S. Wang}



\begin{abstract}
  Deep learning (DL) libraries are widely used in critical applications, where even subtle silent bugs can lead to serious consequences.
  While existing DL fuzzing techniques have made progress in detecting crashes, they inherently struggle to detect silent bugs due to the lack of effective test programs and corresponding oracles.
  
  Building on the observation that historical bug reports contain rich, underutilized information about silent bugs, we leverage large language models (LLMs) to perform \emph{versatile yet controlled} bug transfer for silent bug fuzzing.  
  Specifically, our approach uses LLMs to extract context-aware bug patterns from historical issues, match semantically related Application Programming Interfaces (APIs) using functionality-based embeddings, and synthesize test cases with customized oracles.  
  This enables proactive detection of silent bugs by transferring high-risk contexts and oracle designs from known buggy APIs to functionally similar target APIs.
  To ensure the reliability of our context-aware bug transfer, we introduce an LLM-powered self-validation module that systematically evaluates the validity of each transferred bug instance.  
  We implement this methodology in a tool named \tool and evaluate it on three mainstream DL libraries: PyTorch, TensorFlow, and MindSpore.  
  \tool successfully discovers 79 previously unknown bugs (12 confirmed as Common Vulnerabilities and Exposures (CVEs)) in 10 bug types, demonstrating its effectiveness and generalizability in migrating DL library bug discovery capabilities.

  
\end{abstract}




\maketitle

\section{Introduction}

Deep learning (DL) libraries form the software backbone of many critical
applications, such as autonomous driving, medical diagnosis, and
finance~\cite{liu2020computing, grigorescu2020survey, muhammad2020deep,
wen2023less,li2023commit,wang2023deepvd,li2021vulnerability}. Even subtle bugs
in these libraries can lead to incorrect model behavior with high-stakes
consequences. While crashes are easily observable and memory errors can be
caught by sanitizers, many bugs manifest silently, producing incorrect results
without overt signs of failure~\cite{silent1,silent2}. These silent bugs are
particularly insidious, as they can lead to erroneous outcomes without being
immediately apparent to users or developers. As a result, they are more likely
to propagate through downstream applications, causing significant issues that
may go undetected.

However, detecting silent bugs is inherently challenging due to the lack of general oracles (i.e., mechanisms used to determine whether a bug has been triggered).
Unlike crashes, which exhibit clear and observable symptoms, silent bugs necessitate oracles designed on a case-by-case basis, considering both the specific manifestations of the bug (e.g., incorrect outputs or degraded performance) and the functional characteristics of the involved APIs (e.g., different APIs have different expected outputs)~\cite{silent1,silent2}.
This contextual nature makes it challenging, if not impossible, to predefine oracles for all potential silent bugs.
Existing fuzzers typically rely on rigid, context-insensitive oracles, such as crashes or CPU-GPU result mismatches, which only capture a small subset of silent bugs. As shown in \cref{tbl:comparison}, existing tools~\cite{fuzzgpt,future,dfuzz} collectively detect only 3 of 9 representative types of silent bugs (e.g., performance degradation, wrong outputs).

\begin{wrapfigure}{r}{0.5\textwidth} 
    \centering
    \includegraphics[width=0.48\textwidth]{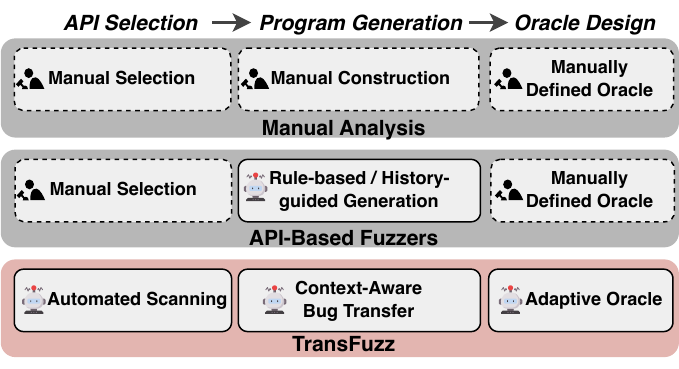}
    \caption{Testing silent bugs: \textit{Manual Analysis} demands hand-crafted test programs for many APIs; existing \textit{API-Based Fuzzers} struggle with rigid API selection and oracle design.}
    \label{fig:intro}
\end{wrapfigure}

Automated testing for silent bugs is challenging due to the dual requirement of generating test programs and designing effective oracles, as illustrated in \F~\ref{fig:intro}. This dual requirement significantly enlarges the testing space, which includes input data, API sequences, and potential oracle behaviors, making exhaustive exploration infeasible. To address this challenge, we should prioritize identifying high-risk APIs and their triggering contexts, focusing testing efforts on these high-stakes scenarios.  To that end, an ideal silent bug detector should achieve three key objectives: (1) identifying high-risk APIs, (2) analyzing triggering contexts, and (3) enabling context-aware oracle design.

Our insight stems from the observation that bugs are not isolated incidents;  
rather, they tend to follow recurring patterns linked to specific functionalities and usage contexts.  
Through quantitative analysis of historical bug reports, we found that APIs with similar functionalities often exhibit analogous bug behaviors in comparable contexts (see \F~\ref{fig:pilotRes} in \S~\ref{sec:pilot}).  
Building on this observation, we propose a proactive strategy: generalize high-risk contexts and oracle designs from known buggy APIs to other APIs with similar functionalities.  
This strategy enables the inference of potentially high-risk APIs and their corresponding triggering contexts from past bug reports, providing actionable guidance for effective oracle design.

Based on the above insight, we present \tool, the first automated testing framework for silent bugs in DL libraries. \tool\ is built on a novel transfer-then-verify pipeline that operates in two stages: bug transfer and bug validation. 
In the transfer stage, \tool extracts context-aware bug patterns from historical issues, including buggy APIs, triggering contexts, and their oracle designs.
These patterns are then transferred to functionally similar target APIs, which are considered potential candidates for similar bugs.
Unlike prior work~\cite{titanfuzz, dfuzz, fuzzgpt}, we employ functionality-based API matching, which computes embeddings of API behavior (rather than surface syntax) to identify appropriate transfer targets.
Once suitable target APIs are identified, large language models (LLMs) are used to synthesize test programs by adapting the context-aware bug pattern to the new API context, automatically generating custom contexts and oracle designs.
While primarily focused on silent bugs, this design can also support regular crash detection.

However, bug transfer alone can introduce false positives, especially when semantic differences between source and target APIs result in behavioral variations.
To mitigate this, \tool introduces a rigorous self-validation phase powered by LLMs.
In this stage, the LLM is prompted to reflect on the original bug report, the transfer rationale, and the runtime outputs of the test program.
It then determines whether the observed behavior constitutes a bug, improving precision for silent bugs.


We implement \tool and evaluate it across three mainstream deep learning frameworks: PyTorch~\cite{pytorch}, TensorFlow~\cite{tf}, and MindSpore~\cite{mindspore}.  
Although originally designed to detect silent bugs, \tool demonstrates strong scalability and can be seamlessly extended to detect crashes without requiring any modifications.  
In our evaluation of PyTorch, \tool successfully identifies 31 silent bugs and 25 crashes.
Notably, most silent bugs are long-lived and hard to expose: 74\% persisted for over three years, and 96\% could not be detected by CPU-GPU differential testing, highlighting the necessity of customized oracles beyond simple differential checks.
To assess its generalizability, we further transfer the bug patterns mined from PyTorch to TensorFlow and MindSpore, uncovering 23 additional crash bugs.
In total, \tool identifies 79 previously unknown bugs in 10 bug types, 12 of which have been assigned CVEs.  
These results confirm the effectiveness of context-aware bug transfer while maintaining broad applicability. 
In sum, we make the following contributions:

\begin{itemize}
    \item We propose a novel \textit{transfer-then-verify} framework that, for the first time, enables automated testing of DL library silent bugs. It systematically extracts context-aware bug patterns and adapts them using functionality-based API matching and LLM-based test synthesis.  

    \item We introduce LLM-powered self-validation that reflects on transfer rationale and test behavior to filter out false positives, enabling reliable detection of silent bugs.

    \item We implement and evaluate \tool across three DL libraries, discovering
    31 silent bugs and 48 crashes (12 CVEs), with high precision on
    silent bugs and strong cross-framework generalizability.
\end{itemize}

\section{Preliminaries and Related Work}
\label{sec:motivation}

In this section, we introduce 
silent bugs and their classification, discuss the challenges faced by existing methods in detecting them, and present our insights and pilot experiments.

\subsection{Crash Bugs vs. Silent Bugs}

Bugs in deep learning (DL) libraries can generally be classified into two categories based on their manifestation: crashes and silent bugs.

\textbf{Crash bugs} are bugs that lead to obvious program failures, such as program crashes, exceptions, infinite hangs, or memory errors. These bugs are typically easier to detect due to their conspicuous symptoms~\cite{wei2022free, titanfuzz, IvySyn, deng2022fuzzing, xie2022docter, yang2023fuzzing}. For instance, exceptions or crashes often produce error messages or core dumps that can be captured and analyzed, infinite hangs can be identified through timeouts, and memory errors can be detected with sanitizers.

\textbf{Silent bugs}, on the other hand, do not produce immediate or obvious symptoms~\cite{silent1,silent2}. Instead, they lead to incorrect program behavior without crashing or raising errors~\cite{xiao2022metamorphic}. These can include incorrect outputs, wrong gradient calculations, or performance degradation. Silent bugs are much harder to detect because they do not manifest through clear failure signals, and are more likely to go unnoticed and persist in downstream applications. Consequently, developing automated techniques to detect silent bugs is a critical yet challenging task.


\noindent\textbf{Classification and Characteristics of Silent Bugs.} 
Building upon existing taxonomies~\cite{silent1}, we classify silent bugs into nine categories based on their manifestation, as summarized in \T~\ref{tbl:comparison}. We first identify eight categories based on distinct observable manifestations, such as discrepancies across execution modes (e.g., \textit{Eager vs.\ Compiled}), hardware platforms (e.g., \textit{CPU vs.\ GPU}), or erroneous gradients and outputs (e.g., \textit{Wrong Gradient}, \textit{Wrong Outputs}). These categories reflect clear and recurring failure patterns.
We use \textit{Functionality Not Working as Expected} to capture the remaining bugs that do not have the above distinctive signatures or occur infrequently.
While our taxonomy does not claim to exhaust all possible forms of silent bugs,
it provides a structured and empirically grounded framework for analyzing and
detecting these silent bugs in deep learning libraries. Importantly, as
will be discussed in \T~\ref{tbl:comparison}, our analysis further reveals that
8 out of 9 silent bug categories lack a general oracle and thus require
context-specific oracle design. Even for the CPU vs GPU category, one of the
most common types, general oracles are only effective for detecting numerical
discrepancies, but not for other unintended behaviors.



This limitation highlights a fundamental challenge in silent bug detection:
silent bugs lack \textbf{universal} correctness criteria, as the definition of
``correct'' outcomes is \textbf{context-dependent}, varying with the specific
API semantics, input data, and API call contexts. As a result, it is infeasible
to define universal oracles that can reliably catch all instances of silent
bugs. Instead, effective detection requires designing customized oracles
tailored to each use case. For example, to identify \textit{Wrong Outputs}, one
must understand the API's semantics and calling contexts, derive the expected
output based on the inputs, and validate the actual output using assertions like
\texttt{torch.allclose()}. Detecting \textit{Wrong Gradients} may involve
numerical gradient checks, while identifying \textit{Eager vs. Compiled}
discrepancies requires side-by-side execution comparisons under controlled
conditions.


\begin{table}[htbp]
\centering
\scriptsize
\caption{Bug detection capabilities by category and tool. Existing tools
  collectively cover only 3 of 9 silent bug types ($\sim$33\%), while \tool
  supports all 9 types.}
\vspace{-5pt}
\setlength{\tabcolsep}{2pt} 
\resizebox{\linewidth}{!}{
\begin{threeparttable}
\begin{tabular}{>{\centering\arraybackslash}m{0.9cm} >{\centering\arraybackslash}m{3cm} >{\centering\arraybackslash}m{1.6cm}
                >{\centering\arraybackslash}m{1.5cm} >{\centering\arraybackslash}m{1.6cm}
                >{\centering\arraybackslash}m{1.5cm} >{\centering\arraybackslash}m{1.5cm}}
\toprule
\textbf{Category} & \textbf{Bug Type} & \textbf{General Oracle Exists?} & \textbf{FUTURE} & \textbf{DFUZZ} & \textbf{FuzzGPT} & \textbf{\tool} \\
\midrule

\multirow{9}{*}{\makecell[c]{Silent\\Bug}}
    & Eager vs Compiled     & Partial\textsuperscript{1}     & \ding{55} & \ding{55} & \ding{55} & \ding{52} \\
    \cline{2-7}
    & Eager vs Just-In-Time (JIT)               & Partial     & \ding{55} & \ding{55} & \ding{55} & \ding{52} \\
    \cline{2-7}
    & CPU vs GPU                 & Partial     & \ding{52} & \ding{52} & \ding{52} & \ding{52} \\
    \cline{2-7}
    & Performance degradation    & No\textsuperscript{2}     & \ding{55} & \ding{55} & \ding{55} & \ding{52} \\
    \cline{2-7}
    & Wrong save/reload         & No     & \ding{55} & \ding{55} & \ding{55} & \ding{52} \\
    \cline{2-7}
    & Wrong displayed message   & No     & \ding{55} & \ding{55} & \ding{55} & \ding{52} \\
    \cline{2-7}
    & Wrong gradient            & Yes     & \ding{55} & \ding{55} & \ding{52} & \ding{52} \\
    \cline{2-7}
    & Wrong outputs             & No     & \ding{55} & \ding{55} & \ding{55} & \ding{52} \\
    \cline{2-7}
    & Functionality Not Working as Expected   & No     & \ding{55} & \ding{55} & \ding{55} & \ding{52} \\
\midrule

Crash & Crash                   & Yes     & \ding{52} & \ding{52} & \ding{52} & \ding{52} \\
\bottomrule
\end{tabular}
\begin{tablenotes}
\small
\item[1] The general oracle can only detect numerical discrepancies, not other unintended behaviors.
\item[2] Each test case requires a different oracle based on its specific context.
\end{tablenotes}
\end{threeparttable}
}
\label{tbl:comparison}
\vspace{-10pt}
\end{table}

\subsection{Existing Works and Their Limitations}
\label{sec:related}

Based on the characteristics of silent bugs, the ideal automated testing for silent bugs in DL library APIs should have the following key features:
\begin{itemize}
    \item Be able to identify APIs prone to silent bugs.
    \item Be able to analyze context information that may trigger these bugs, including API calling context (e.g., specific API call sequences or interactions with other APIs) and input data.
    \item Be able to design an oracle based on the context information to detect potential bugs.
\end{itemize}

Despite recent advances, current testing techniques still fall short of meeting these requirements when detecting silent bugs. Specifically, they face the following four challenges:

\noindent\textbf{Challenge I: Inefficient Identification of High-Risk APIs.}
Current methods struggle to effectively pinpoint APIs susceptible to silent bugs.
This inefficiency largely stems from two common but limited strategies:  
(1) \textit{Rule-based static matching}. Tools like DFUZZ~\cite{dfuzz} and FUTURE~\cite{future} use manually defined rules to choose target APIs. For example, DFUZZ matches APIs based on parameter type similarity, and FUTURE uses cross-framework mappings. While these methods work in some cases, they rely on manual rules and are ineffective for silent bugs.  
(2) \textit{Random matching}. For example, FuzzGPT~\cite{fuzzgpt} randomly selects target APIs, resulting in low testing efficiency and poor coverage of high-risk APIs.

\noindent\textbf{Challenge II: Inefficient Identification of High-Risk Contexts.}  
When testing silent bugs in a target API, both the input space (e.g., data types, values, shapes, attributes) and the interaction space with other APIs must be considered. Each of these spaces is vast, and combining them results in an even larger search space. Methods like random mutation, used by tools such as TitanFuzz and DFUZZ, cannot fully explore this space. Some researchers have used historical bug reports to guide test program generation strategies (e.g., FuzzGPT, FUTURE).
However, they typically generate new test cases by embedding the original bug code into prompts.
This approach often leads to test programs that only match superficial syntactic patterns of the original bug code, without deeply investigating the root causes of the bug.
As a result, the generated tests tend to ``overfit'' the original bug code.
For example, when handling trigger conditions, these methods typically copy the original conditions directly, without adapting them to the characteristics of the target API.

\noindent\textbf{Challenge III: Difficulty in Designing Context-Aware Oracles.}
Existing DL library fuzzers typically rely on two main oracle strategies, both of which have significant limitations in detecting context-dependent silent bugs.
The first observes coarse-grained runtime signals, such as error messages, crashes, or infinite hangs~\cite{titanfuzz, xie2022docter, gu2022muffin}.  
The second uses pre-defined oracles for specific types of silent bugs. For instance, some fuzzers include a \textit{CPU vs. GPU} consistency check to detect backend-related discrepancies, while \textit{FuzzGPT} additionally validates gradient correctness.



However, many other types of silent bugs like \textit{Wrong Displayed Message} and \textit{Functionality Not Working as Expected} cannot be captured using such pre-defined oracles.
We assess three state-of-the-art fuzzing tools (FuzzGPT, DFUZZ, and FUTURE) in terms of their coverage of nine silent bug categories.
As shown in \T~\ref{tbl:comparison}, even when combining all three methods, only 3 out of 9 scenarios (i.e., $\sim$33\%) are covered, leaving 6 scenarios unexplored.

Moreover, manually designing oracles based on the context of each case is impractical, as variations in input data and expected behaviors would require different oracles. For example, with 100,000 test cases, we cannot predict which types of silent bugs may exist in each case. To cover all nine bug types, we would need to design 900,000 unique oracles, which is clearly unfeasible for real-world testing. This gap highlights the need for context-aware oracles that can dynamically detect bugs based on specific API characteristics, input data, and bug manifestations.

\noindent\textbf{Challenge IV: High False Positive Rate.}
Although some specific types of silent bugs already have expert-driven, general differential testing oracles, such as CPU vs. GPU detection, these methods are not entirely reliable.
Output inconsistencies do not necessarily indicate the presence of a bug.
Such differences can stem from inherent program behavior, such as the use of random functions in input construction, where outputs naturally vary between runs.
Consequently, approaches like TitanFuzz~\cite{titanfuzz} and DFUZZ generate a large number of false positives in practice, requiring significant manual effort to analyze and verify each case.

\begin{figure*}[h]
\centering
\includegraphics[width=\linewidth]{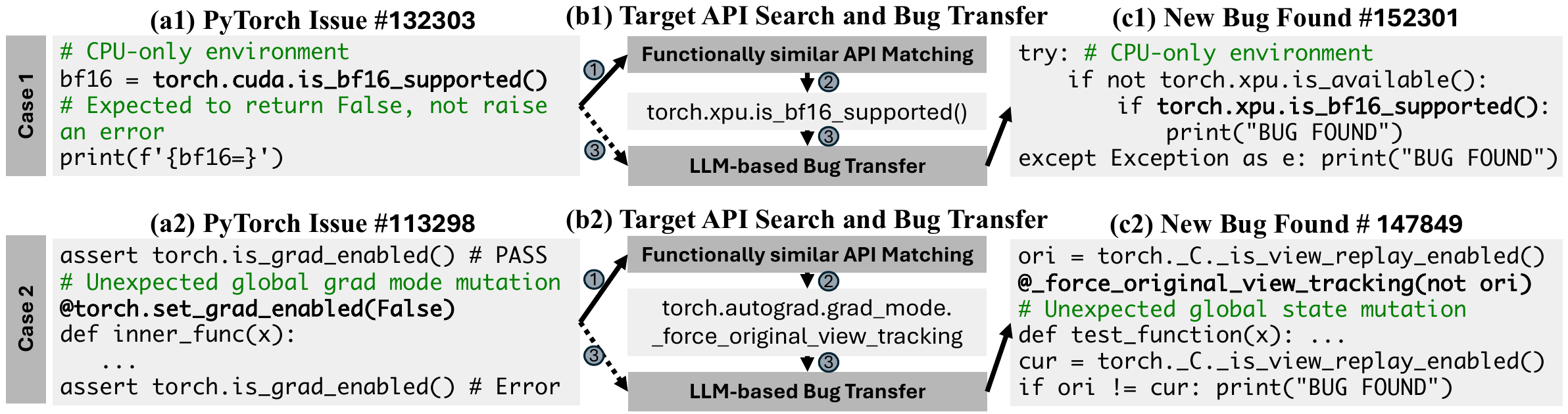}
\vspace{-5pt}
\caption{Two Examples of Silent Bug Transfers.}
\vspace{-15pt}
\label{fig:motivation2}
\end{figure*}


\subsection{Motivating Insights and Examples}
The limitations outlined above arise from a core shortcoming: existing methods treat bug discovery as isolated, API-agnostic tasks, without systematically leveraging historical knowledge or cross-API semantic patterns to guide testing and oracle design. Our key innovation lies not merely in applying LLMs to fuzzing workflows, but in designing a \textit{pattern-transfer-driven} framework that bridges known bugs with potentially vulnerable APIs through structured reasoning.
Our approach is grounded in two key insights:
\textit{Insight 1}: Historical bug reports contain a wealth of unexploited information about silent bugs.
\textit{Insight 2}: API bugs, especially in API implementations, are rarely isolated. When APIs share similar functionality, they often follow comparable design and implementation patterns, leading to analogous vulnerabilities. For instance, numerical computation errors may recur across functionally similar APIs when handling analogous boundary conditions. In \S~\ref{sec:pilot}, we further conduct a quantitative analysis to validate this observation.

Based on Insight 1, we can extract known bug APIs, their triggering contexts, and oracle designs from historical bug reports. 
Drawing from Insight 2, we can identify candidate target APIs that are functionally similar to the known buggy API. These candidates may suffer from analogous silent bugs (\textit{Challenge I}).
We then instruct the LLM to generate custom triggering contexts and oracle designs based on the characteristics of the target API, integrate them into a test program, and execute the program to verify whether similar bugs exist (\textit{Challenges II \& III}).
Finally, to reduce false positives (\textit{Challenge IV}), we employ the LLM to reflect on the entire transfer process, evaluating whether the inferred bug pattern is logically and semantically plausible in the new context.
By transferring known bug patterns across similar APIs with LLM-powered adaptation and validation, we form a closed-loop pipeline that systematically uncovers silent bugs.

\noindent\textbf{Motivating Example.}
As shown in \F~\ref{fig:motivation2}, we illustrate our approach using two representative \emph{original} silent bug issues, which serve as source bug patterns for transfer.
Case 1 corresponds to an environment-related logic bug in \texttt{issue \#132303}, where \texttt{torch.cuda.is\_bf16\_supported} raises an exception instead of returning \texttt{False} when executed in a CPU-only environment.
Case 2 corresponds to a global state mutation bug reported in \texttt{issue \#113298}, in which certain context-management APIs unexpectedly modify global execution states (e.g., gradient mode), leading to silent semantic inconsistencies.

\begin{itemize}
    \item \textbf{Addressing Challenge I (High-Risk API Identification):} 
    Instead of random testing, we transfer bug patterns extracted from original issues to other functionally similar APIs to identify high-risk testing targets.
    For Case 1, we match \texttt{torch.xpu.is\_bf16\_supported} as a target API because it performs analogous hardware capability checks under different device backends.
    For Case 2, we identify \texttt{\_force\_original\_view\_tracking} as a target, as it serves a similar role in managing global execution states.

    \item \textbf{Addressing Challenges II \& III (Context \& Oracle Design):}
    After identifying target APIs for bug pattern transfer, we further transfer
    and adapt the associated triggering contexts and oracle designs from the
    original issues. Existing fuzzers such as TitanFuzz and FreeFuzz rely on
    general-purpose oracles (e.g., crashes or CPU-GPU output consistency),
    which cannot detect these logic bugs, as they exhibit
    fully consistent behavior across CPU and GPU executions.
    Identifying such logic bugs typically requires developers to manually design
    tailored triggering contexts and oracles for each case, which is
    labor-intensive and difficult to scale. In Case 1, our tool
    extracts specific triggering contexts (e.g., ``CPU-only environment'') and
    custom oracles (e.g., ``Check if return is \texttt{True}'' or ``Catch specific
    warning messages'') from historical issue and related patches and adapts them
    to the target API. In Case 2, it captures the implicit assumption
    that context managers should not mutate global states, and designs an oracle
    that checks whether execution states (i.e., view-replay mode) remain
    unchanged before and after API invocation. Consequently, an oracle trigger
    serves as a diagnostic signal for a potential bug.

    \item \textbf{Addressing Challenge IV (High False Positives):} 
    To ensure the correctness of transferred patterns, we apply an LLM-powered self-validation step.
    For example, in Case 2, our tool reasons about whether a detected global state change reflects a true semantic violation or an intended design choice.
    Only behaviors that are inconsistent with documented semantics are reported as bugs.
    Detailed mechanisms and additional examples are presented in \S~\ref{sec:verify}.
\end{itemize}

\subsection{Pilot Experiments}
\label{sec:pilot}

To investigate our hypothesis that bugs in DL libraries exhibit recurring patterns tied to specific API functionalities and usage contexts, we conduct a pilot study analyzing real-world bug reports from PyTorch.
Our goal is to assess whether similarities in API functionality and triggering context correlate with similarities in how bugs manifest, particularly in terms of oracle design.

\begin{figure}[h]
\centering
\includegraphics[width=0.5\linewidth]{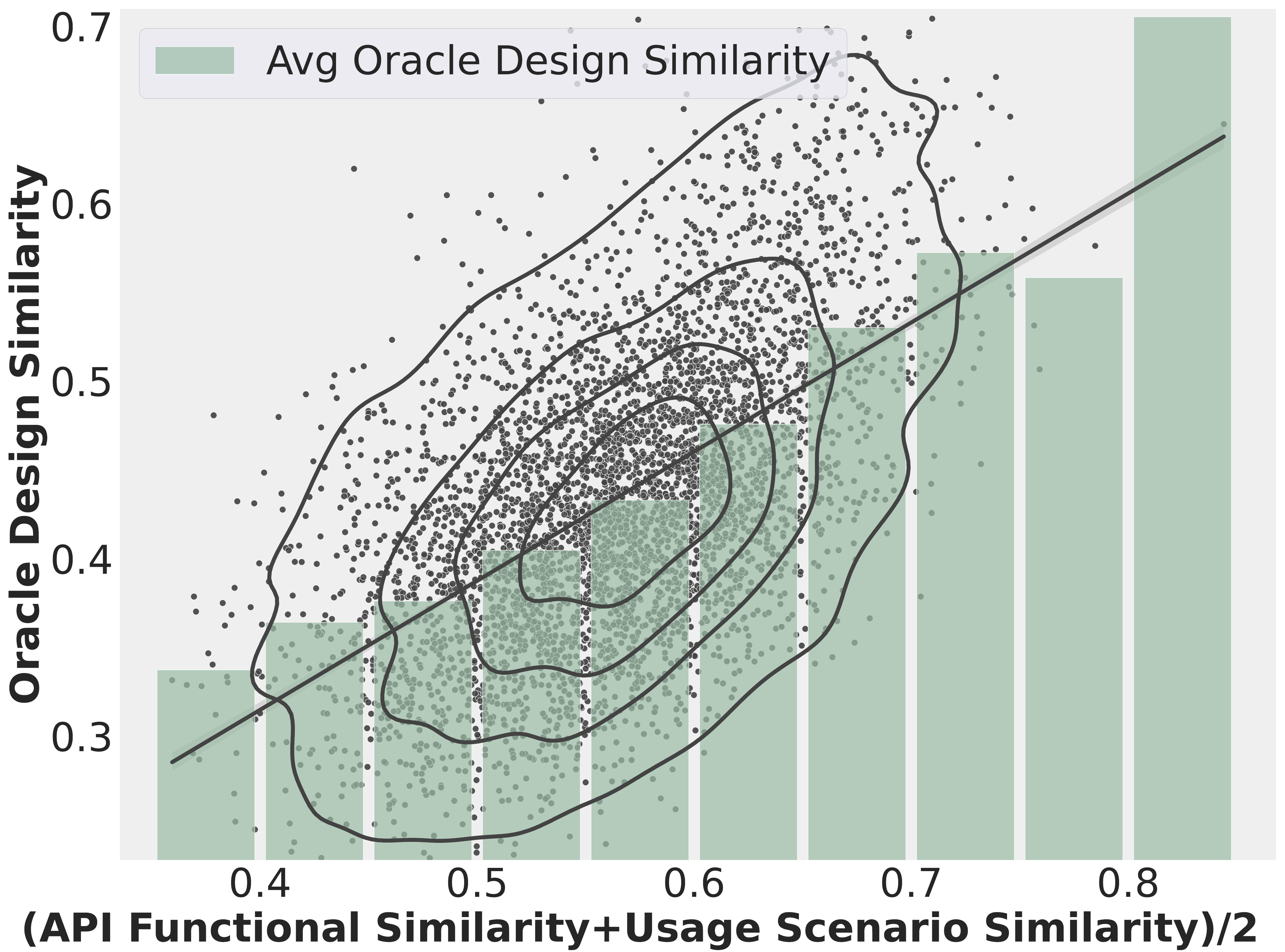}
\vspace{-5pt}
\caption{Oracle Design Similarity vs. Functional and Usage Scenario Similarity.}
\vspace{-10pt}
\label{fig:pilotRes}
\end{figure}

\noindent\textbf{Data Collection.}
We collect all GitHub issues labeled as ``correctness (silent)'' from the official PyTorch repository, which indicate silent correctness errors.
After removing irrelevant discussions (e.g., feature requests), we randomly sample 100 bug reports for in-depth analysis.

\noindent\textbf{Information Extraction.}
For each selected issue, we use GPT-4.1-mini to extract three key attributes: (1) \textit{bug API functional description}, (2) \textit{triggering context}, and (3) \textit{oracle design}. These attributes are then manually verified, and embedded into vectors using the \texttt{text-embedding-3-small} model.
As a result, each of the 100 issues is represented by a triplet of embedding vectors: (API functional description, usage scenario, oracle design).


\noindent\textbf{Similarity Analysis and Visualization.} 
We compute the cosine similarity between every pair of issues across all three dimensions, yielding 9,900 pairwise comparisons.
Each comparison includes three similarity scores: API functional, triggering context, and oracle design similarity.
To visualize these relationships, we map each pair into a 2D space: the x-axis is the average of API functional and usage scenario similarities \textit{(x = 0.5 $\times$ API functional similarity + 0.5 $\times$ usage scenario similarity)}, and the y-axis is the \textit{oracle design similarity}. \F~\ref{fig:pilotRes} shows the resulting scatter plot, along with regression lines, density contours, and bar charts that display the mean oracle design similarity within 0.05-wide x-axis bins.

\noindent\textbf{Observation.}  
The plots reveal a clear trend: \textit{when two issues involve APIs with similar functionalities and triggering contexts, they are more likely to share similar bug manifestations and oracle designs.} We further quantified this using Pearson correlation analysis, yielding a Pearson correlation coefficient of $r = 0.561$ ($r > 0$ signifies a positive association) and a $p$-value of $0.000$ ($p < 0.05$ denotes statistical significance). This result confirms a statistically significant positive correlation between API functionality/context and bug manifestation.

\section{Design}

\begin{figure*}[h]
    \centering
    \vspace{-10pt}
    \includegraphics[width=\linewidth]{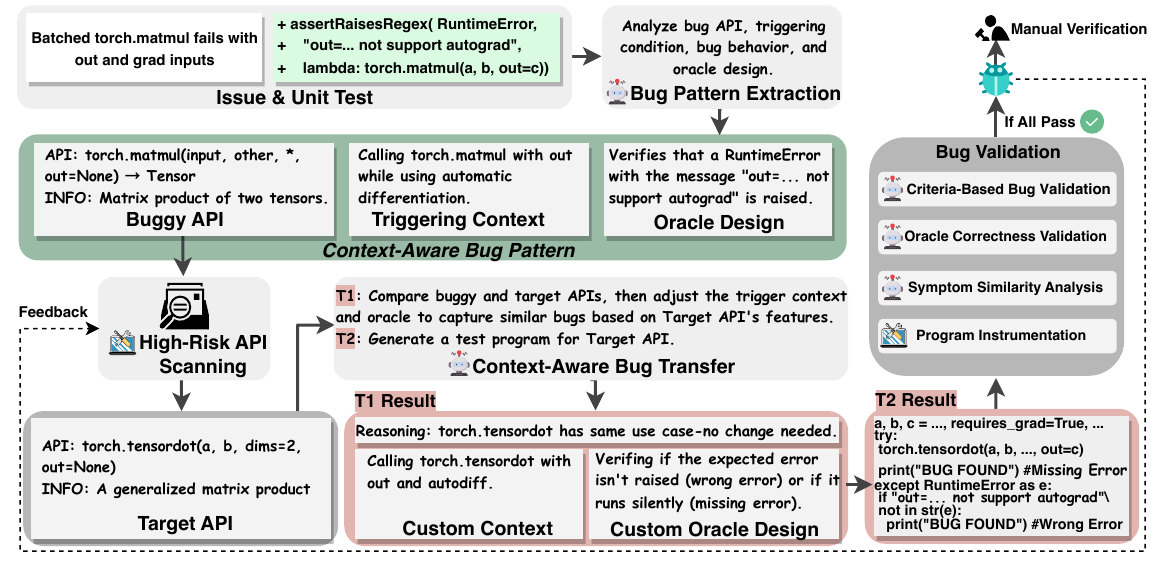}
    \vspace{-15pt}
    \caption{The Workflow of \tool, which consists of four main stages: Bug Pattern Extraction, High-Risk API Scanning, Context-Aware Bug Transfer, and Bug Validation.}
    \vspace{-10pt}
    \label{fig:overview}
\end{figure*}


\subsection{Overview}
To address the challenges highlighted in \S~\ref{sec:motivation}, we present \tool, a novel transfer-then-verify framework that fuzzes silent bugs through \emph{versatile yet controlled} bug transfer.
\F~\ref{fig:overview} shows the workflow of \tool. It consists of four key components:

\noindent
\ding{172}
\textit{Context-Aware Bug Pattern Extraction}: \tool begins by mining historical bug reports. For each bug, it extracts a \textit{context-aware bug pattern} that encapsulates three critical elements: the buggy API, the triggering context, and the oracle design. These patterns serve as structured templates to guide subsequent bug transfer.

\noindent
\ding{173} 
\textit{Functionality-Based API Matching}: 
To generalize bug patterns beyond the original API, \tool identifies functionally similar APIs by computing semantic similarity between context-free functional descriptions of the original API and other library APIs. This produces a prioritized queue of candidate APIs (i.e., the similar API queue) likely to exhibit similar vulnerabilities.

\noindent
\ding{174}
\textit{Bug Transfer-Driven Fuzzing}:  
Given a bug pattern and the similar API queue, \tool generates targeted test programs.
It selects target APIs using a feedback-driven strategy based on the similar API queue, enabling efficient identification of APIs likely to exhibit similar bugs.
For each target API, \tool uses LLMs to adapt the bug pattern by rewriting the triggering context and redesigning the oracle to match the new API's semantics. This yields customized, executable test programs.

\noindent
\ding{175}
\textit{LLM-Powered Self-Validation}:  
Since LLM-generated oracles are not always reliable, \tool performs post-execution validation using LLMs to minimize false positives. By reflecting on the original issue, transfer rationale, and runtime behavior of the generated tests, \tool significantly improves the reliability and precision of bug detection.

\subsection{Context-Aware Bug Pattern Extraction}
\label{sec:pattern}

\noindent\textbf{Context-Aware Bug Pattern.} 
Extracting known bug information from historical bug reports forms the foundation of effective bug transfer.
The quality of this extraction directly impacts the success of the transfer process.
We introduce the concept of \textit{context-aware bug pattern}, which encapsulates all critical elements involved in triggering and detecting silent bugs, including
(i) \textit{Bug API}: The specific API where the bug was originally discovered;
(ii) \textit{Triggering Context}: The conditions or actions that activate the bug, such as invalid inputs or a particular sequence of API calls;
and
(iii) \textit{Oracle Design}: An automated mechanism for detecting the bug in code, enabling identification without manual inspection.

\noindent\textbf{Why is it necessary to extract context-aware bug patterns? }
A straightforward approach to bug transfer is to directly use historical bug code to guide test program generation, as seen in tools like FuzzGPT and FUTURE.
However, this method may overlook that bug reports often contain irrelevant details and contextual noise, such as unrelated variables or redundant control structures.
If we prompt LLMs to generate test programs directly from such noisy reports, aiming at triggering similar bugs in target APIs, the models may struggle to abstract the core bug logic.
Instead, they tend to overfit on superficial syntactic elements, failing to capture and transfer the essential characteristics of the bug.
As a result, the generated test programs could be often ineffective.

To address this, we propose context-aware bug patterns as an intermediate representation in the bug transfer pipeline.
Instead of directly transferring from a bug report to a test case, we adopt a staged approach: 
\textit{Original Bug Report $\rightarrow$ Context-Aware Bug Pattern $\rightarrow$ New Bug in the Target API}.
This enhances the accuracy and generalizability of bug transfer.
By first distilling a precise bug pattern, we enable the LLM to generate a customized triggering context and oracle that align with the unique characteristics of the target API, ultimately improving the likelihood of uncovering potential similar bugs.

\noindent\textbf{Our Method.}
In this paper, we use PyTorch as an example to demonstrate our approach.\footnote{We also transfer bug patterns from PyTorch to the other two DL libraries for crash detection; see \S~\ref{sec:bugDiscovery}.}
To enable effective bug transfer, we first collect real-world bug instances that are both reproducible and user-relevant.
Specifically, we crawl issues and pull requests (PRs) from the PyTorch repository, focusing on those that reflect bugs observable from the user's perspective. 
Given that PyTorch has over 50,000 issues, analyzing each one would consume an excessive number of LLM tokens.
We thus focus only on issues that have corresponding PRs, ensuring that the issue is a confirmed bug that has been addressed and fixed.
In the future, we could leverage LLMs to analyze the issues further, such as identifying whether the bugs are user-triggerable, and thereby extract additional bug patterns.

Building on this curated dataset, we then leverage the advanced code comprehension capabilities of LLMs~\cite{devlin2018bert, feng2020codebert, brownlanguage, yadavally2023partial, li2023contextuality} to extract context-aware bug patterns from both the issue discussions and the associated code changes, using a carefully designed prompt to guide systematic analysis.
As illustrated in \F~\ref{fig:issueAnalysis} in \cref{sec:prompts}, this prompt guides the LLM to analyze issue reports and code changes, identify the exact API involved, clarify the conditions required to trigger the bug, contrast the expected versus actual behaviors, and pinpoint the oracle used to detect the bug.
Moreover, the prompt further directs the model to generate a user-facing Python test program that can reproduce the bug, ensuring that the extracted patterns are actionable and grounded in real-world usage scenarios.

\subsection{Functionality-Based API Matching}
\label{sec:matching}

After extracting the context-aware bug pattern from the original issue, our next step is to identify candidate transfer targets, which are other APIs that might exhibit similar bugs. 
Our analysis of historical bug reports reveals a key empirical pattern: APIs with similar functionalities tend to exhibit analogous bugs under comparable contexts; see \F~\ref{fig:pilotRes} in \S~\ref{sec:pilot}.
Leveraging this insight, we use functional similarity as a guiding principle to identify candidate transfer targets. 
Unlike DFUZZ, which performs transfer mainly at the \emph{input level} based on syntactic properties (e.g., input types and signatures), \tool performs transfer at the \emph{API (functional) level} by modeling semantic functionality, enabling the transfer of bug-triggering logic and oracle designs even across APIs with different signatures. This broader transfer capability is reflected in our evaluation: \T~\ref{tbl:comparison} shows that DFUZZ covers only 1 of 9 silent bug categories, whereas \tool supports all 9, and \T~\ref{tab:compare} further indicates that DFUZZ detects only one silent bug, compared to 31 detected by \tool.



To implement this matching process, we start by automatically scanning the target library to collect all available APIs, including their function names and module paths.
We then crawl documentation and parameter details for each API.
Next, leveraging well-crafted prompts (see \F~\ref{fig:apiAnalysis} in Appendix~\ref{sec:prompts}), we employ LLMs to distill these documents, generating context-free functional descriptions that capture input parameter types and core operation workflows.
Once we have these descriptions, we encode them into fixed-dimension vectors using embedding models like the OpenAI embedding API, such as \textit{text-embedding-3-small}.
All embeddings are stored systematically within an API embedding database for easy retrieval.
Finally, using the API embedding database, we compute cosine similarity scores for each context-aware bug pattern, identifying the top 1,000 APIs most similar to the original bug API.
This forms the \textit{similar API queue}, which is used for subsequent test case generation.


\subsection{Bug Transfer-Driven Fuzzing}
\label{sec:transferFuzzing}
Having laid the groundwork in previous sections, we now present our core methodology of bug transfer-driven fuzzing.
Specifically,
in \S~\ref{sec:pattern}, we extract context-aware bug patterns for each issue, including details such as the original bug API. 
In \S~\ref{sec:matching}, we generate a similar API queue for each original bug API based on context-free functional similarity.
We integrate these components into a fuzzing framework that performs bug transfer-guided test generation.

\noindent\textbf{Overview.}
Our bug transfer-driven fuzzing consists of two components: \textit{Feedback-Driven API Selection} and \textit{Context-Aware Bug Transfer}.
Leveraging the similar API queue, we propose a feedback-driven API selection strategy to efficiently identify target APIs likely to exhibit similar bugs.
For each target API, we design context-aware triggering context and oracles based on API functionality to detect analogous bugs and automatically generate test programs for validation.

\A~\ref{alg:fuzzing} outlines the entire fuzzing process.
Given an issue and its fix, our method first extracts a context-aware bug pattern via LLMs (\S~\ref{sec:pattern}) and builds a queue of similar APIs (\S~\ref{sec:matching}) in lines 2-3.
The fuzzing loop iteratively selects up to 10 untested APIs from the queue (line 6), and for each newly discovered buggy API, adds its top-10 similar APIs to expand the search (lines 7-9).
For each target API, the system generates a test by transferring the bug pattern and runs it (lines 10-12).
If a bug is found and verified, the API is recorded (lines 13-15).
The process continues until no new bugs are detected, enabling iterative, feedback-driven bug discovery.

\begin{algorithm}
\caption{Bug Transfer-Driven Fuzzing.}
\label{alg:fuzzing}
\footnotesize

\KwIn{An issue and its corresponding PR}
init := true\\
$\text{bug\_pattern} \gets$ \texttt{LLM\_extract(issue, PR)}\\
$\text{api\_queue} \gets$ \texttt{API\_match(bug\_pattern.api)}\\
$\text{found\_new\_bug\_api} \gets \emptyset$\\

\While{$\text{init}$ \textbf{or} $\text{found\_new\_bug\_api} \neq \emptyset$}{
    $\text{targets} \gets$ top-10 untested APIs from \texttt{api\_queue}\\
    \ForEach{$a \in \text{found\_new\_bug\_api}$}{
        add top-10 similar APIs of $a$ to \texttt{targets}\\
        remove $a$ from \texttt{found\_new\_bug\_api}\\
    }
    \ForEach{$a \in \text{targets}$}{
        $p \gets$ \texttt{Transfer(bug\_pattern, $a$)}\\
        \texttt{run($p$)}\\
        \If{\texttt{BUG FOUND}}{
            \If{\texttt{self\_validation(issue, $p$)}}{
                add $a$ to \texttt{found\_new\_bug\_api}\\
            }
        }
    }
    \texttt{init := false}\\
}
\end{algorithm}

\noindent\textbf{Feedback-Driven API Selection.}  
Under limited testing resources, we propose a feedback-driven API selection strategy to efficiently identify APIs likely to harbor similar bugs based on the similar API queue.  
We iteratively conduct batch testing following the similar API queue, as shown in \F~\ref{fig:overview}.  
In each round, we select the top N untested APIs (where N is the window size, set to 10 based on the results in \S~\ref{sec:hyperparameter}) with the highest similarity scores for bug transfer.  
If any bugs are detected in a round, the next batch proceeds; otherwise, the process terminates early.  
This strategy allows us to quickly focus on high-risk areas while avoiding wasted resources on low-yield candidates.  

When a transferred test program for a target API uncovers a new bug, we classify the functionality of that API as high-risk.  
We then use the embedding of this target API as an anchor to identify and test the 10 most similar APIs.  
This feedback loop enables us to go beyond the static ranking of the initial queue and adaptively explore APIs that are more likely to contain similar bugs, guided by actual fuzzing outcomes.

\noindent\textbf{Context-Aware Bug Transfer.}
To enable effective migration of the original bug to target APIs, we introduce a context-aware bug transfer approach.
\F~\ref{fig:bugTransfer} (in Appendix~\ref{sec:prompts}) illustrates the prompt that completes this entire synthesis process in a Chain-of-Thought (CoT) manner. Specifically:

First, we analyze the functional semantics and usage scenarios of both the original bug API and the target API.
This comparison highlights their differences, providing a robust semantic foundation for subsequent bug transfer reasoning.

Next, assuming that the target API triggers similar bugs, we prompt the LLM to infer its potential triggering context and abnormal behavior, such as returning error codes, throwing exceptions, or causing resource leaks.
Importantly, even for similar bugs, the precise triggering contexts often differ across APIs.
For instance, an edge-case bug in API A may be triggered when \textit{input = 0}, while the equivalent edge case in API B may instead occur at \textit{input = -1}.

Then, based on the expected abnormal behavior, \tool generates a corresponding oracle to catch it.
Oracle design is guided by both API characteristics and bug types, requiring careful consideration of the specific functional properties of the API and the nature of the bug.
For example, oracle designs for ``wrong displayed message'' and ``wrong gradient'' are fundamentally different.
Typically, the oracles include assertions on return values, checks for exception types and error messages, and resource state confirmations, thereby ensuring clear test verdicts and strong coverage.

Finally, \tool synthesizes complete test programs by integrating the triggering context, API invocation sequence, and oracle design.
These programs are executable out-of-the-box and serve to validate whether the hypothesized bug is indeed reproducible in the target API.

\subsection{LLM-Powered Self-Validation}
\label{sec:verify}

After completing the bug transfer process, we generate the corresponding test
programs tailored to the target APIs. When executed, if a test program outputs
\textit{``BUG FOUND''}, indicating that the oracle has been triggered, it
suggests the presence of a potential bug. However, due to the inherent
randomness and variability in LLM-generated code, the oracles themselves may
contain inaccuracies or over-generalizations, leading to false positives. As a
result, not all triggered oracles reflect real bugs in the target API.
Therefore, further validation is required to confirm whether they are false
positives.


\noindent\textbf{Insight.}
In theory, if comprehensive and precise documentation clearly defined the expected behavior of an API under all conditions, LLMs would be able to identify false positives.
However, the real-world use of APIs is highly complex, often involving numerous edge cases.
In practice, documentation typically provides only a brief overview of API functionality and descriptions of input and output parameters.
As a result, directly relying on LLMs to determine whether a case is a false positive often leads to a high rate of misclassification.
As shown in \T~\ref{tab:datasetRes} (see Appendix~\ref{sec:ablationVer}), we also explore the direct use of LLMs in the curated dataset to determine whether a transferred test program triggers a real bug.
Experimental results show that the precision of this approach is only 30.77\%.
It is important to note that 28.20\% of the samples in this dataset are real bugs.
This means that even a naive strategy that predicts every sample as a real bug would achieve a precision of 28.20\%.

To reduce false positives, it is crucial to establish reliable criteria for judgment.
We observe that fixed bugs serve as ideal reference points, since they have been identified by developers.
By extracting the \textbf{bug determination criteria} (i.e., the core rationale behind labeling specific program behaviors as bugs) from these resolved cases, we can guide automated reasoning.
More importantly, by establishing these criteria, we transform the subjective
task of determining whether a test triggers a real bug into an objective
classification with clear decision grounds. This novel approach mitigates LLM
hallucinations and forms the foundation of our validation process. Leveraging
the bug determination criteria allows us to analyze the runtime behavior of
transferred tests to determine whether they are false positives.

\begin{figure}[h]
    \centering
    \vspace{-10pt}
    \includegraphics[width=0.7\linewidth]{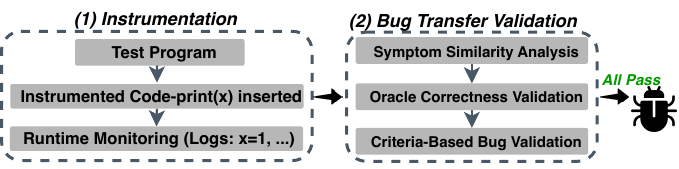}
    \vspace{-5pt}
    \caption{Overview of LLM-powered self-validation.}
    \vspace{-10pt}
    \label{fig:verifyOverview}
    \end{figure}

\noindent\textbf{Overview.}  
Based on the above insight, we propose an \textit{LLM-powered self-validation} approach that automatically filters false positives without any human involvement.
\F~\ref{fig:verifyOverview} illustrates the overall workflow, which consists of two tightly coupled components:
\textit{\textbf{(1) Instrumentation}} and \textit{\textbf{(2) Bug Transfer Validation}}.

Given a transferred test program that triggers an oracle (i.e., outputs \textit{``BUG FOUND''}), directly trusting this result is unreliable.
In practice, an oracle may be triggered by unintended program behaviors rather than the intended manifestation of a real bug, due to corner cases, environment-specific effects, or deficiencies in the oracle design.
To disambiguate these scenarios, we therefore examine the program's \emph{actual runtime behavior}.
To this end, we introduce automated \textit{Instrumentation} to capture fine-grained runtime states during execution.
Instrumentation produces execution logs that record the program's concrete runtime behavior, such as relevant variable states.
These logs serve as the factual input to the subsequent \textit{Bug Transfer Validation}, grounding the LLM's reasoning in concrete execution evidence.

Our \textit{Bug Transfer Validation} consists of three complementary checks to determine whether the transferred program triggers a real bug:
\textit{(i) Symptom Similarity Analysis}: Given the functional similarity between the original and target APIs, comparable behavior is expected when similar bugs are triggered. Thus, we assess the behavior similarity between the original issue and the transferred program from multiple perspectives. \textit{(ii) Oracle Correctness Validation}: Assuming the target API is bug-free, we verify that the oracle is logically valid and not triggered by environmental noise or LLM-induced hallucinations. \textit{(iii) Criteria-Based Bug Validation}: Extracting the bug determination criteria from the original issue and using them to assess whether the transferred program is a false positive.
Only when all checks are satisfied do we conclude that the transferred program triggers a real bug.

\noindent
\textbf{1) Instrumentation.}  
The role of instrumentation is to bridge the gap between oracle-level signals and semantic bug reasoning.
While an oracle only reports that a bug condition is met, instrumentation reveals how the program actually reaches that condition at runtime.
This includes the executed path, the specific oracle that is triggered, and the runtime states that lead to the trigger.
Such execution evidence enables the LLM to determine whether an oracle trigger reflects the intended manifestation of a real bug or results from unintended side effects.
Therefore, instrumentation is applied uniformly across all subsequent validation stages. Specifically, the execution logs generated by the instrumentation are leveraged in the prompts for Symptom Similarity Analysis (\F~\ref{fig:sameBugPattern}), Oracle Correctness Validation (\F~\ref{fig:bugFree}), and Criteria-Based Bug Validation (\F~\ref{fig:insightCheck}).

We implement instrumentation via automated Abstract Syntax Tree (AST) transformation on the transferred test program.
The transformation is lightweight, fully automated, and preserves original program semantics.
It supports the following language constructs:

\textit{Variable Assignments}: We cover ordinary assignments (\textit{a = 1}), attribute assignments (\textit{obj.attr = 1}), index assignments (\textit{lst[0] = 2}), and multi-variable unpacking (\textit{a, b = 1, 2}). For each variable's first assignment, we insert a print statement (e.g., \textit{print(``{var} ='', {var})}). 

\textit{Exception Handling}: For exception variables captured via \textit{except ... as e}, a print statement is injected at the entry of the exception handler to log the exception state.

\textit{Context Managers}: For context manager variables in \textit{with ... as w} statements, we place instrumentation at the beginning of the controlled block to monitor the runtime value of the acquired resource.

\textit{Conditional Expressions}: We implement recursive parsing for complex nested structures within \textit{if} conditions, instrumenting only those sub-expressions that can be independently evaluated. 
For instance, in the condition \textit{if torch.isnan(x.grad).any()}, instrumentation is applied to \textit{x.grad}, \textit{torch.isnan(x.grad)}, and the complete condition itself, ensuring that only expressions with actual runtime values are captured. Additionally, our method supports extraction across multi-level call chains. For expressions such as \textit{a.b().c[0].d()}, instrumentation is applied stepwise to \textit{a.b()}, \textit{a.b().c}, \textit{a.b().c[0]}, and finally \textit{a.b().c[0].d()}. 

\begin{table}[h]
    \setlength{\tabcolsep}{2pt}
    \centering
    \vspace{-5pt}
    \caption{Key IR identifiers and their meanings.}
    \vspace{-5pt}
    \footnotesize
    \begin{tabular}{@{}p{3.2cm}p{5cm}@{}}
    \toprule
    \textbf{IR Identifier} & \textbf{Description} \\
    \midrule
    \texttt{VarDef(\textless type\textgreater)} & 
    Defines a variable of given type, e.g., \texttt{tensor}. \\
    \texttt{APICall(\textless api\textgreater)} & 
    Represents a function call and its return value. \\
    \texttt{OracleCheck(\textless type\textgreater)} & 
    Specifies assertions, e.g., \emph{ValueCorrectness}. \\
    \texttt{condition=} & 
    Specifies comparison logic, e.g., \texttt{Compare(v1,v2)}. \\
    
    \bottomrule
    \end{tabular}    
    \vspace{-5pt}
    \label{tab:ir_identifiers}
  \end{table}

\noindent
\textbf{2) Symptom Similarity Analysis.}
Since our fuzzing framework is based on migrating bug patterns across functionally similar APIs, it is reasonable to expect a high degree of behavioral consistency between the original issue and its transferred programs.
Accordingly, our validation focuses on whether the transferred program preserves the essential bug pattern of the original issue under a different API context.
To this end, we organize the analysis as a two-step validation pipeline that progressively filters incorrect transfers while avoiding unnecessary fine-grained analysis.
We first check whether the original issue and the transferred case belong to the same bug type, which serves as a coarse-grained prerequisite for further validation.
Based on the identified bug type, we then apply different checks:
for mismatch bugs, we verify whether the transferred program triggers a real mismatch;
for all other bug types, we examine whether the two cases exhibit the same bug pattern by comparing their trigger conditions, runtime behaviors, and oracles.

We begin by determining whether the original issue and the transferred test program belong to the same bug type.
A discrepancy at this stage signals an incorrect bug transfer and the case is immediately filtered.
Directly prompting LLMs to compare bug types often leads to inconsistent or self-justifying outputs.
To make this process objective and reproducible, we introduce an intermediate representation (IR) that formally defines seven common bug types.
Key IR identifiers are summarized in \T~\ref{tab:ir_identifiers}.
This IR-based formulation transforms subjective semantic comparison into structured pattern matching, effectively reducing LLM hallucination.
The prompt used for this classification is shown in \F~\ref{fig:sameBugType} in Appendix~\ref{sec:prompts}.
As an example, the IR definition of a device inconsistency bug is as follows:
\[
\begin{array}{l}
\text{Device\_Inconsistency} ::= \\
\quad \texttt{VarDef(tensor)[device=cpu:*]} \rightarrow v_{cpu} \wedge \\
\quad \texttt{VarDef(tensor)[device=cuda:*]} \rightarrow v_{gpu} \wedge \\
\quad \texttt{OracleCheck(ValueCorrectness)}( \\
\quad\quad\ \texttt{condition=Compare}(v_{cpu}, v_{gpu}) \\
\quad) \rightarrow \texttt{FAIL}
\end{array}
\]
Other bug types are defined in a similar manner in Appendix~\ref{sec:bugIR}.

For mismatch bugs, the oracle is typically simple and reliable, as it only checks for output inconsistencies of the same program across different environments.
As a result, we do not perform additional fine-grained bug pattern analysis, but only need to double-check whether this is a mismatch bug based on the prompt in \F~\ref{fig:realMismatch}.
Otherwise, for all non-mismatch bug types, oracle reliability is often lower and a stricter semantic comparison is required.
In such cases, we assess whether the original issue and the transferred test program share the same bug pattern from three complementary perspectives: bug trigger conditions, runtime behaviors, and oracle designs.
Leveraging LLMs together with instrumented execution logs, we analyze the consistency of these dimensions between the source and transferred cases.
The prompt used for this analysis is shown in \F~\ref{fig:sameBugPattern}.

\begin{figure}[h]
\centering
\includegraphics[width=\linewidth]{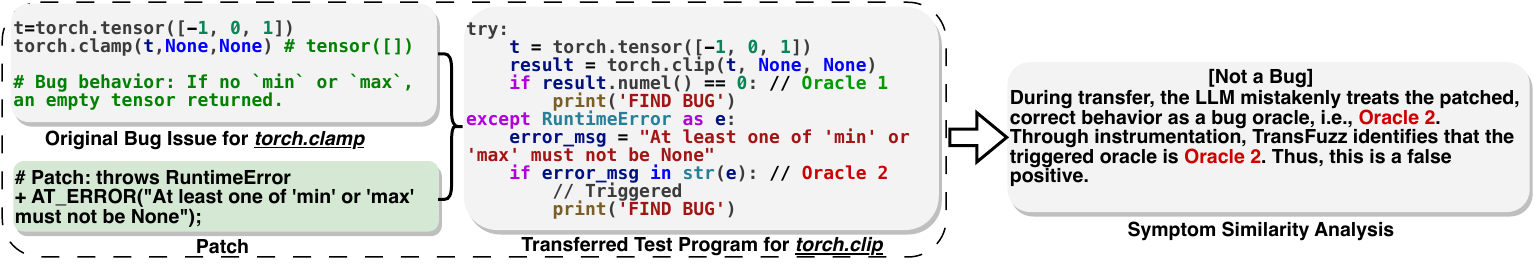}
\caption{False positive filtered by symptom similarity analysis. The
  transferred program for \texttt{torch.clip} incorrectly checks error messages
  instead of empty output, revealing a semantic mismatch from the original
  \texttt{torch.clamp} bug.}
\label{fig:example1}
\end{figure}

\F~\ref{fig:example1} illustrates how symptom similarity analysis filters out a false positive caused by an incorrectly transferred oracle.
First, at the code level, the original \textit{torch.clamp} issue is classified as a \texttt{Functional\_Defect}, as its oracle checks whether an empty tensor is returned when both \textit{min} and \textit{max} are \textit{None}.
The transferred program for \textit{torch.clip} similarly validates functional correctness by checking the runtime outcome, either by inspecting whether the output is empty or whether the raised error message contains a target string.
Since both oracles reason about functional behavior, the transferred program passes the coarse-grained bug-type consistency check.

\tool then conducts a finer-grained analysis using instrumented execution logs to assess whether the original and transferred programs exhibit the same bug pattern.
The analysis uncovers a clear discrepancy: the original oracle characterizes the bug by detecting an empty output tensor, whereas the oracle actually triggered in the transferred program checks whether the raised error message matches a specific string.
As these two oracles encode fundamentally different bug patterns, \tool identifies the transferred case as a false positive and filters it out.
Further inspection reveals that the error-message-based oracle in the transferred program is incorrect, as it is introduced by the LLM mistakenly treating the patched, correct behavior of \textit{torch.clip} (raising a \textit{RuntimeError}) as a bug oracle.

\noindent
\textbf{3) Oracle Correctness Validation.}
For the cases that pass the checks in symptom similarity analysis,
we use a reverse hypothesis validation method to analyze the correctness of the oracle design, specifically to determine whether it might be triggered incorrectly.
We begin by assuming that the target API does not contain a bug similar to the one in the original issue. Under this assumption, if the oracle in the transferred test case still gets triggered, this implies a flaw in the oracle's design, as it should not flag a problem in the absence of a bug.
The prompt for this analysis is shown in \F~\ref{fig:bugFree} in Appendix~\ref{sec:prompts}.
As shown in \F~\ref{fig:example2}, we use an example to demonstrate how oracle correctness validation identifies false positives induced by invalid semantic assumptions in the transferred oracle. The original bug originated from \textit{torch.mul} violating the commutative property under the \textit{torch.compile} mode, where the oracle specifically detects whether $f(a, b) \neq f(b, a)$. When this bug pattern is transferred to \textit{torch.fmod}, the test similarly flags non-commutative behavior as a bug. However, \textit{fmod(x, y)} and \textit{fmod(y, x)} are mathematically non-commutative by definition, meaning the oracle can be triggered even when the implementation is fully correct. Oracle correctness validation identifies that the oracle encodes an invalid semantic assumption for the transferred API and therefore filters this case as a false positive.

\begin{figure}[h]
\centering
\includegraphics[width=\linewidth]{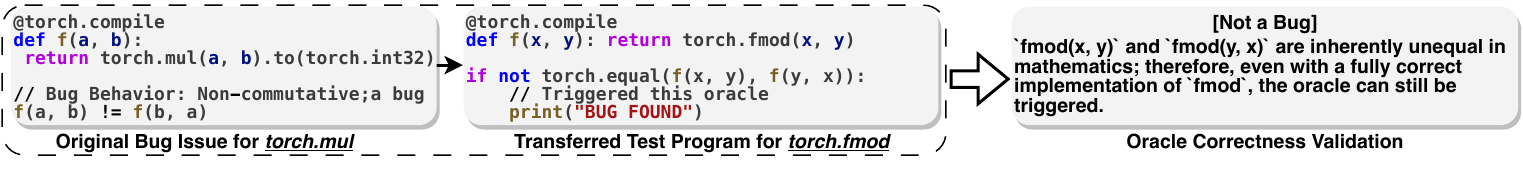}
\caption{False positive filtered by oracle correctness validation. The
   non-commutative property holds for \texttt{torch.mul} but not for
   \texttt{torch.fmod} by mathematical definition.}
\label{fig:example2}
\end{figure}

\noindent
\textbf{4) Criteria-Based Bug Validation.}  
For the cases that pass the checks in oracle correctness validation,
we use a criteria-based bug validation method to ultimately determine whether this is a real bug.
The core of this analysis lies in extracting bug determination criteria from the original issue, then using these criteria to assess whether the transferred test program genuinely triggers a bug in the target API. To achieve this, the process unfolds in three steps.

First, assess whether reliable bug determination criteria can be extracted from the original issue by evaluating four dimensions: API-level bug relevance, presence of a demo, developer's negative feedback, and bug complexity, with prompts shown in \F~\ref{fig:issueCheck1}, \F~\ref{fig:issueCheck2}, \F~\ref{fig:issueCheck3}, and \F~\ref{fig:issueCheck4} in Appendix~\ref{sec:prompts}, respectively. An issue is considered suitable for extraction only if it satisfies all four checks.

Next, analyze the original issue and, based on the functional characteristics of the API, justify why the observed behavior should be classified as an API bug rather than expected behavior. This involves identifying the specific functional requirement or intended behavior that is violated, forming the foundational bug determination criteria. The prompt guiding this analysis is shown in \F~\ref{fig:insightAnalysis} in Appendix~\ref{sec:prompts}.

\begin{figure}[h]
\centering
\includegraphics[width=\linewidth]{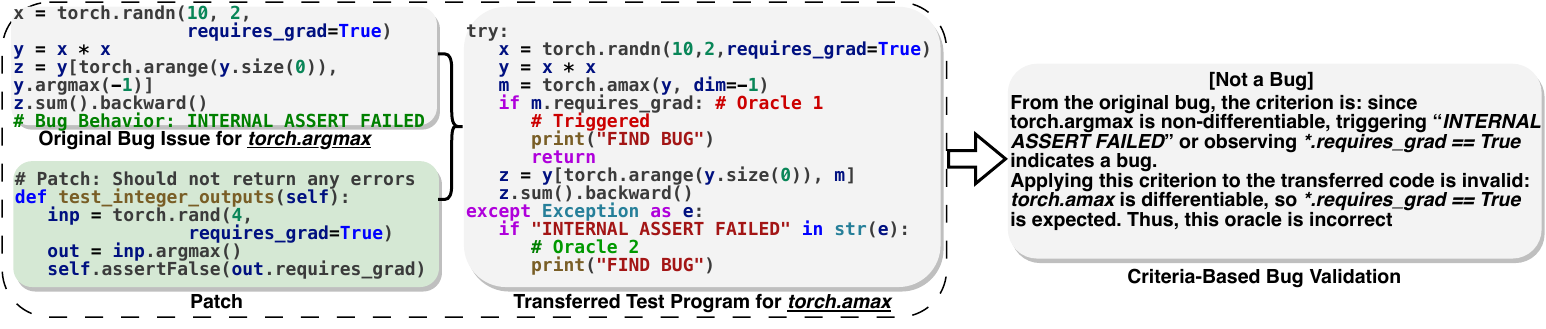}
\caption{False positive filtered by criteria-based bug validation.
  \texttt{torch.argmax} is non-differentiable by design, but
  \texttt{torch.amax} is differentiable.}
\vspace{-5pt}
\label{fig:example3}
\end{figure}

Finally, based on the execution outputs, analyze the transferred test program to determine why the oracle is triggered. Specifically, why \textit{``BUG FOUND''} is printed. Then, using the bug determination criteria extracted from the original issue as a reasoning framework, evaluate whether the observed behavior in the target API is a real bug. This involves assessing whether the oracle's trigger reflects a real violation of expected functionality, rather than an inconsistency in test design or environmental factors. The prompt guiding this analysis is provided in \F~\ref{fig:insightCheck} in Appendix~\ref{sec:prompts}. To enhance accuracy, we introduce a debate mechanism. After receiving the initial prompt response in \F~\ref{fig:insightCheck}, the LLM will adopt a counterpoint perspective, questioning the original interpretation (i.e., ``Please challenge this from the opposing viewpoint''). Ultimately, the final assessment will summarize whether this constitutes a false positive, based on both perspectives (i.e., ``Based on both viewpoints, summarize whether this is a false positive'').
As shown in \F~\ref{fig:example3}, we present an illustrative example of how criteria-based bug validation identifies and filters false positives by verifying the applicability of the original bug's logic to the transferred API. From the original \textit{torch.argmax} issue, we extract a bug criterion stating that since \textit{argmax} is non-differentiable, triggering an \textit{INTERNAL ASSERT FAILED} error or observing \textit{requires\_grad = True} signals a bug. This criterion is then applied in a transferred test for torch.amax, with the oracle checking whether the output tensor propagates gradients. However, unlike \textit{argmax}, \textit{amax} is differentiable by design, and \textit{requires\_grad $=$ True} is therefore expected behavior rather than a bug symptom. Criteria-based bug validation detects that the transferred case violates the semantic preconditions underlying the original criterion and correctly filters it as a false positive.

\section{Implementation and Evaluation}

The core testing process of \tool is implemented in Python, consisting of
about 3,600 lines of code. 
Our implementation provides automated issue
analysis, API analysis, and test generation to support our evaluation.
To evaluate the effectiveness of \tool, we focus on the following research questions (RQs):
\begin{itemize} 
  \item \textbf{RQ1}: How many previously unknown bugs and distinct types of bugs can \tool discover?
  \item \textbf{RQ2}: How does \tool compare with existing tools in terms of effectiveness?
  \item \textbf{RQ3}: How many detected bugs are real?
  \item \textbf{RQ4}: How are the hyperparameters determined?
  \item \textbf{RQ5}: What are the results of the manual analysis of prompt effectiveness?
  \item \textbf{RQ6}: What is the cost of using \tool?
\end{itemize}

\subsection{Experimental Setup}

\noindent\textbf{Environment.}
All experiments are performed on a machine equipped with 256 GB of RAM, an NVIDIA RTX A6000 GPU, and Ubuntu 20.04 as the operating system.


\noindent\textbf{LLMs.} We use o3-mini~\cite{o3mini} for \textit{Context-Aware Bug Pattern Extraction}. Given that generating context-free functional descriptions in \textit{Functionality-Based API Matching} is relatively straightforward, we choose to use GPT-4o mini~\cite{gpt4omini}. Since \textit{Bug Transfer-Guided Test Generation} requires generating a large number of test cases, we adopt GPT-4o mini for efficiency. For \textit{LLM-Powered Self-Validation}, we utilize GPT-4.1 mini~\cite{gpt41} for its stronger reasoning capabilities.


\begin{table}[t]
\caption{Silent bugs discovered in PyTorch by \tool and their corresponding bug transfer paths. Underlined issue IDs (e.g., \underline{torch \#152299}) indicate bugs involving API interactions. Note: A single issue may contain multiple bugs.}
\label{tab:silent}
\centering
\resizebox{\linewidth}{!}{
\begin{tabular}{l|l|l|l|c}
\toprule
\textbf{Bug Type} & \textbf{Bug API} & \textbf{Source Issue} & \textbf{Bug Found} & \textbf{Num} \\
\midrule
\makecell[l]{CPU VS GPU}
& torch.sparse.log\_softmax & torch \#38839 & torch \#152293 & 1 \\
\midrule
\makecell[l]{Eager VS Compiled}
& torch.Tensor.asinh\_ + torch.compile & torch \#118267 & \underline{torch \#152299} & 1 \\
\midrule
\makecell[l]{Eager VS Jit}
& torch.set\_default\_dtype + torch.jit.script & torch \#36369 & \underline{torch \#150607} &  1\\
\midrule
\makecell[l]{Performance Degradation}
& torch.functional.pca\_lowrank & torch \#86234 & torch \#153534 &  1\\
\midrule
\makecell[l]{Wrong Save/Reload}
& torch.ao.nn.quantized.Sigmoid & torch \#91877 & torch \#147818 &  1\\
\midrule
\multirow{2}{*}{Wrong Displayed Message}
& torch.utils.data.default\_collate & torch \#47160 & torch \#153536 & \multirow{7}{*}{7} \\
& torch.fft.ihfft2 + torch.fft.rfft2 & torch \#67140 & \underline{torch \#149625} &  \\
& torch.fft.ihfftn + torch.fft.rfftn & torch \#59127 & \underline{torch \#149625} &  \\
& torch.randn + torch.compile & torch \#147070 & \underline{torch \#149625} &  \\
& torch.Tensor.sum + torch.compile & torch \#96484 & \underline{torch \#149625} &  \\
& torch.onnx.operators.reshape\_from\_tensor\_shape + torch.compile & torch \#94831 & \underline{torch \#149625} &  \\
& torch.rand\_like + torch.jit.trace & torch \#128581 & \underline{torch \#147844} &  \\
\midrule
\multirow{2}{*}{\shortstack{Functionality Not Working \\as Expected}}
& torch.tensordot & torch \#117067 & torch \#147846 & \multirow{14}{*}{14} \\
& torch.arange + torch.vmap & torch \#102208 & \underline{torch \#152295} &  \\
& torch.ones + torch.vmap & torch \#102208 & \underline{torch \#152295} &  \\
& torch.scalar\_tensor + torch.vmap & torch \#102208 & \underline{torch \#152295} &  \\
& torch.utils.\_sympy.functions.make\_opaque\_bitwise\_fn & torch \#138395 & torch \#147841 &  \\
& torch.jit.trace\_module & torch \#52217 & torch \#154478 &  \\
& torch.utils.data.default\_collate + torch.set\_default\_dtype & torch \#36369 & \underline{torch \#150607} &  \\
& torch.cuda.manual\_seed + torch.compile & torch \#107187 & \underline{torch \#149621} &  \\
& torch.cuda.manual\_seed\_all + torch.compile & torch \#107187 & \underline{torch \#149621} &  \\
& torch.cuda.random.manual\_seed + torch.compile & torch \#107187 & \underline{torch \#149621} &  \\
& torch.xpu.random.set\_rng\_state\_all + torch.compile & torch \#107187 & \underline{torch \#149621} &  \\
& torch.xpu.random.manual\_seed\_all + torch.compile & torch \#107187 & \underline{torch \#149621} &  \\
& torch.xpu.manual\_seed\_all + torch.compile & torch \#107187 & \underline{torch \#149621} &  \\
& torch.autograd.grad\_mode.\_force\_original\_view\_tracking & torch \#113359 & torch \#147849 &  \\
\midrule
\makecell[l]{Wrong Gradient}
& torch.nn.functional.hardswish & torch \#51438 & torch \#147801 &  1\\
\midrule
\multirow{2}{*}{Wrong Outputs}
& torch.addcmul & torch \#98691 & torch \#152294 & \multirow{4}{*}{4} \\
& torch.add & torch \#98691 & torch \#152294 &  \\
& torch.xpu.is\_bf16\_supported & torch \#132303 & torch \#152301 &  \\
& torch.\_dynamo.utils.is\_compile\_supported & torch \#111179 & torch \#147826 &  \\
\bottomrule
\end{tabular}
}
\vspace{-10pt}
\end{table}

\subsection{RQ1: Bug Discovery}
\label{sec:bugDiscovery}

Although \tool is originally designed to detect silent bugs, it exhibits strong scalability and can be seamlessly extended to detect crashes without requiring any modifications. We evaluate its effectiveness in both scenarios across three mainstream deep learning frameworks: PyTorch~\cite{pytorch}, TensorFlow~\cite{tf}, and MindSpore~\cite{mindspore}.

\noindent\textbf{Silent Bug Detection in PyTorch.} 
We conducted silent bug detection on PyTorch versions 2.6.0 and 2.7.0, utilizing 7,466 historical issues as source data.
\T~\ref{tab:silent} summarizes the results of silent bug detection in PyTorch using \tool. A total of 31 silent bugs were identified across nine categories, with the majority belonging to three primary types: Functionality Not Working as Expected (14 bugs), Wrong Displayed Message (7 bugs), and Wrong Outputs (4 bugs). Each bug was identified by transferring patterns from a known source issue and subsequently reported as a new issue. Notably, some source issues triggered multiple bugs, indicating the presence of reusable bug patterns. Each of the remaining categories (i.e., CPU vs GPU, Eager vs Compiled, Eager vs JIT, Performance Degradation, Wrong Save/Reload, Wrong Gradient, and Wrong Outputs) accounted for a single instance. These results demonstrate the effectiveness of \tool in uncovering a diverse range of silent bugs through context-aware pattern mining.

To evaluate the complexity of silent bugs found by \tool, we analyzed whether they involve API interactions.
The results are shown in \T~\ref{tab:silent}, where bugs with underlined IDs (e.g., \underline{torch \#152299}) indicate that their triggering process involves API interactions.
While existing tools mostly detect simple, standalone API calls, over 50\% of the silent bugs identified by \tool involve inter-API interactions, indicating its ability to uncover state-dependent bugs.

The 31 silent bugs found by \tool expose critical risks to PyTorch's real-world reliability. Their consequences fall into four high-impact categories:
(i) \textit{Model Failure}: Gradient or output errors silently corrupt training/inference (e.g., \texttt{hardswish} grad, torch\#147801; \texttt{addcmul} outputs, torch\#152294).
(ii) \textit{Deployment Breakage}: Mode/hardware mismatches cause silent divergence in production (e.g., CPU/GPU \texttt{log\_softmax}, torch\#152293; Eager/Compiled \texttt{asinh\_}, torch\#152299).
(iii) \textit{Production Reliability Loss}: Serialization failures, misleading errors, or state corruption break deployment pipelines (e.g., quantized \texttt{Sigmoid} reload, torch\#147818; seed APIs under \texttt{compile}, torch\#149621).
(iv) \textit{Performance Degradation}: Critical slowdowns undermine scalability (e.g., \texttt{pca\_lowrank}, torch\#153534).

In addition, among the 31 silent bugs, 23 (74\%) persisted for over three
years, and 30 (96\%) could not be detected by CPU-GPU differential testing,
demonstrating the inherent difficulty of exposing such bugs and establishing the
necessity of customized oracles beyond simple differential
checks.

\noindent\textbf{Crash Detection and Cross-Framework Transfer.}  
Encouraged by its success in silent bug detection, we applied \tool to crash detection, a key metric in fuzzing~\cite{g2fuzz, truzz, shapfuzz, liu2023vd, she2024fox, xie2025hfuzz, li2025enhancing, xiao2023phyfu}. 
\T~\ref{tab:crash} shows the results.  
\tool successfully uncovered 25 crash bugs from diverse sources, validating its broad applicability.  
To evaluate generalizability, we transferred the context-aware bug patterns mined from PyTorch to TensorFlow (v2.18.0) and MindSpore (v2.3.0).  
Silent bug detection requires precise, context-dependent oracles.
Due to substantial behavioral divergence across frameworks (e.g., error handling, exception semantics, and execution models), reliably transferring such oracles across frameworks is fundamentally challenging.
As the first work to enable fully automated silent-bug testing, we therefore focus on intra-library transfer, where semantic alignment makes oracle construction and validation feasible.
For cross-framework settings, we instead consider crash bugs, whose manifestations are more explicit and stable.
Furthermore, \tool was not specifically designed for crash detection. 
Differences in the crash-triggering conditions between PyTorch and TensorFlow APIs may have caused the bug patterns extracted from PyTorch to be less effective when applied to TensorFlow.

  \begin{table}[htbp]
  \caption{Crash bugs discovered by \tool with corresponding bug transfers.}
  \label{tab:crash}
  \centering
  \resizebox{\linewidth}{!}{
  \begin{tabular}{l|l|l|c||l|l|l|c}
  \toprule
  \textbf{Library} & \textbf{Origin Issue} & \textbf{New Bug Report} & \textbf{Num} &
  \textbf{Library} & \textbf{Origin Issue} & \textbf{New Bug Report} & \textbf{Num} \\
  \midrule
  \multirow{25}{*}{PyTorch} 
    & torch \#98857 & torch \#154425 & \multirow{25}{*}{25} &
  \multirow{19}{*}{\shortstack{MindSpore}} 
    & torch \#117252 & mindspore \#IBVKM8 & \multirow{19}{*}{19} \\
    & torch \#63583 & torch \#154424 &  &
    & torch \#117252 & mindspore \#IBVKM8 &  \\
    & torch \#63583 & torch \#154424 &  &
    & torch \#120803 & mindspore \#IBWMCY &  \\
    & torch \#45967 & torch \#154423 &  &
    & torch \#10654  & mindspore \#IBWMCY &  \\
    & torch \#21833 & torch \#154422 &  &
    & torch \#21526  & mindspore \#IBWMCY &  \\
    & torch \#113141 & torch \#154420 &  &
    & torch \#84170  & mindspore \#IBWMCY &  \\
    & torch \#113141 & torch \#154420 &  &
    & torch \#71629  & mindspore \#IBWMCY &  \\
    & torch \#105934 & torch \#154419 &  &
    & torch \#118379 & mindspore \#IBWMCY &  \\
    & torch \#39708 & torch \#150836 &  &
    & torch \#74438  & mindspore \#IBWMCY &  \\
    & torch \#118990 & torch \#149800 &  &
    & torch \#21526  & mindspore \#IBWMCY &  \\
    & torch \#35977 & torch \#149737 &  &
    & torch \#28214  & mindspore \#IBWMCY &  \\
    & torch \#29367 & torch \#149737 &  &
    & torch \#122171 & mindspore \#IBWMCY &  \\
    & torch \#47608 & torch \#149724 &  &
    & torch \#102208 & mindspore \#IBWMCY &  \\
    & torch \#56920 & torch \#149724 &  &
    & torch \#1605   & mindspore \#IBWMCY &  \\
    & torch \#72106 & torch \#149724 &  &
    & torch \#84144  & mindspore \#IBWMCY &  \\
    & torch \#143484 & torch \#149626 &  &
    & torch \#17750  & mindspore \#IBWMCY &  \\
    & torch \#45967 & torch \#149623 &  &
    & torch \#105311 & mindspore \#IBWMCY &  \\
    & torch \#20529 & torch \#149622 &  &
    & torch \#75568  & mindspore \#IBWMCY &  \\
    & torch \#20529 & torch \#149622 &  &
    & torch \#15918  & mindspore \#IBWMCY &  \\
    & torch \#85237 & torch \#147722 &  & & & & \\
    & torch \#31472 & torch \#150835 &  &
  \multirow{4}{*}{TensorFlow} 
    & torch \#4570 & tf \#90910 & \multirow{4}{*}{4} \\
    & torch \#69688 & torch \#149821 &  & & torch \#4737 & tf \#90241 &  \\
    & torch \#63583 & torch \#149820 &  & & tf \#58270 & tf \#90915 &  \\
    & torch \#24731 & torch \#149822 &  & & tf \#58270 & tf \#90915 &  \\
    & torch \#140923 & torch \#149274 &  & & & & \\
  \bottomrule
  \end{tabular}%
  }
  \end{table}

In total, \tool discovers 79 new bugs across the three frameworks, 12 of which have been confirmed as CVEs (see Table~\ref{tab:cve} in Appendix), with Common Vulnerability Scoring System (CVSS) scores up to 4.8 and the majority classified as MEDIUM severity. Except for CVE-2025-2149, which is a silent bug caused by incorrect save/reload behavior, all others are crash bugs (e.g., segmentation faults, floating point exceptions). As summarized in \T~\ref{tab:silent} and \T~\ref{tab:crash}, these results demonstrate that \tool is not only effective in detecting silent bugs within PyTorch but also highly scalable in crash detection. We further summarize the oracle types used by \tool (see \F~\ref{fig:oracleCount} in Appendix~\ref{sec:oracleCount}), highlighting its ability to automatically design diverse oracles based on bug behavior. 
We also present an analysis of API transfer relationships, summarized in Appendix~\ref{sec:apiRelationship}, which reveals that \tool can trigger a diverse set of bug transfers beyond simple parallel APIs by exploiting shared semantic invariants.

\begin{figure}[h]
  \centering
  \includegraphics[width=0.55\linewidth]{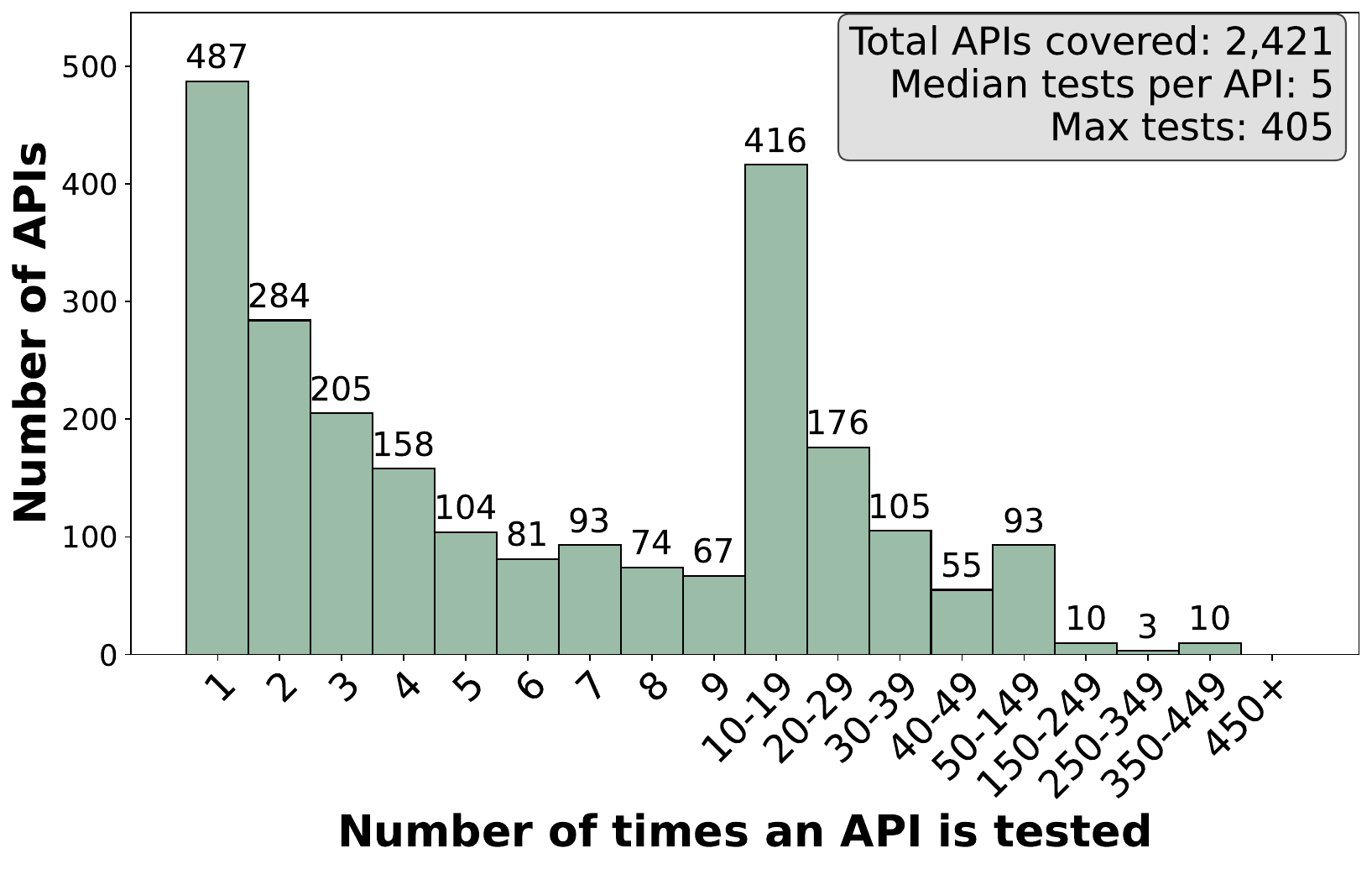}
  \vspace{-10pt}
  \caption{Frequency distribution of APIs tested by \tool in PyTorch.}
  \vspace{-10pt}
  \label{fig:applicability}
\end{figure}

\noindent\textbf{Applicability Analysis.}
To assess whether \tool's approach is broadly applicable (i.e., effective across a wide range of APIs rather than limited to a few), we analyzed the range of APIs it can target in PyTorch. Unlike tools like TitanFuzz, which require users to manually specify target APIs, \tool automatically selects APIs as fuzzing targets by matching functionally similar APIs to those with historical bugs. A key concern is whether this automated strategy restricts the scope of APIs targeted.
As shown in Figure~\ref{fig:applicability}, we evaluated the number of APIs \tool selected for testing and the distribution of their test frequencies. \tool targeted 2,421 APIs in PyTorch for fuzzing, compared to 1,593 APIs targeted by TitanFuzz, DFUZZ, FuzzGPT (based on user-provided lists). This demonstrates that our method can effectively target a broad range of APIs for testing.


\subsection{RQ2: Comparison with State-of-the-Art Fuzzers}
\label{sec:otherFuzzers}

\noindent\textbf{Baseline Fuzzer Selection.} We evaluate \tool against six
fuzzers: FreeFuzz~\cite{wei2022free}, DeepREL~\cite{deng2022fuzzing},
ACETest~\cite{acetest}, TitanFuzz~\cite{titanfuzz}, DFUZZ~\cite{dfuzz}, and
Pathfinder~\cite{pathfinder}. We selected FreeFuzz, DeepREL, TitanFuzz, and
DFUZZ because they represent the state-of-the-art in API-level fuzzing tools.
Among these, TitanFuzz~\cite{titanfuzz} and DFUZZ~\cite{dfuzz} are LLM-based
fuzzers, while Pathfinder~\cite{pathfinder} and ACETest~\cite{acetest} are
specialized fuzzers targeting kernel functions within deep learning libraries.
We excluded the following fuzzers from our evaluation: (1)
FuzzGPT~\cite{fuzzgpt}: It is not open-source. (2) FUTURE~\cite{future}: It
lacks adaptation for PyTorch. (3) IvySyn~\cite{IvySyn}: A critical
dependency is no longer
available, making it impossible to conduct experiments.\footnote{\url{https://github.com/driazati/breakpad}}

\noindent\textbf{Experimental Setup.}
To ensure a fair comparison, we try to maintain a consistent set of target APIs across all fuzzers. Specifically, we incorporate all relevant APIs listed in \T~\ref{tab:silent} and \T~\ref{tab:crash} into each fuzzer's target scope. However, some tools are not fully open-sourced, which limits their extensibility. For instance, FreeFuzz and DeepREL do not provide their constraint extraction or API matching algorithms; consequently, we are unable to adapt them to test certain APIs.

We primarily evaluate all tools on the latest PyTorch v2.6. However, as Pathfinder is only compatible with PyTorch v2.2, we perform a separate head-to-head comparison between \tool and Pathfinder on the older v2.2 environment to ensure a fair assessment. The results are summarized in \T~\ref{tab:compare}.

\noindent\textbf{Results.} The results in \T~\ref{tab:compare} demonstrate that
\tool significantly outperforms existing tools in both bug count and the
diversity of bug types.
\begin{itemize}
  \item Superior Crash Detection: On PyTorch v2.6, \tool identified 25 crashes,
  which is 3.5$\times$ more than the best-performing baseline, ACETest (7
  crashes). Other fuzzers like FreeFuzz and TitanFuzz failed to detect any
  crashes. When compared with Pathfinder on PyTorch v2.2, \tool remains highly
  competitive. While Pathfinder found 24 crashes (slightly more than \tool's
  22), it remains a specialized tool focused solely on crash detection.
  \item Unique Multi-type Bug Discovery: Unlike baseline fuzzers which are
  largely confined to detecting crashes or simple CPU/GPU inconsistencies, \tool
  successfully uncovered a wide array of silent bugs. These include 14
  functional errors, 7 wrong error message bugs, and critical cross-mode
  inconsistencies (e.g., Eager vs. Compiled/JIT), most of which are beyond the
  reach of traditional fuzzing oracles.
\end{itemize}

\begin{table}[h]
  \caption{Comparison of \tool and existing tools in terms of bug count and type diversity.}
  \vspace{-5pt}
  \label{tab:compare}
  \centering
  \resizebox{\linewidth}{!}{
  \begin{threeparttable}
  \begin{tabular}{l|cccccc|cc}
  \toprule
  \multirow{2}{*}{\textbf{Bug Type}} & \multicolumn{6}{c|}{\textbf{Tested on Torch 2.6}} & \multicolumn{2}{c}{\textbf{Tested on Torch 2.2}} \\
  \cmidrule(lr){2-7} \cmidrule(lr){8-9}
  & \textbf{FreeFuzz} & \textbf{DeepREL} & \textbf{ACETest} & \textbf{TitanFuzz}  & \textbf{DFuzz}  & \textbf{\tool} & \textbf{Pathfinder} & \textbf{\tool} \\
  \midrule
  Crash & 0 & 6 & 7 & 0 & 8 & \textbf{25} & \textbf{24} & 22 \\
  \midrule
  CPU VS GPU & 1 & N/A\textsuperscript{*} & N/A & 1 & 1 & \textbf{1} & N/A & \textbf{1} \\
  \midrule
  Eager VS Compiled & N/A & N/A & N/A & N/A & N/A  & \textbf{1} & N/A & \textbf{1} \\
  \midrule
  Eager VS JIT & N/A & N/A & N/A & N/A & N/A  & \textbf{1} & N/A & \textbf{1} \\
  \midrule
  Performance Degradation & N/A & N/A & N/A & N/A & N/A  & \textbf{1} & N/A & \textbf{1} \\
  \midrule
  Wrong Save/Reload & N/A & N/A & N/A & N/A & N/A  & \textbf{1} & N/A & \textbf{1} \\
  \midrule
  Wrong Displayed Message & N/A & N/A & N/A & N/A & N/A  & \textbf{7} & N/A & \textbf{7} \\
  \midrule
  \makecell[l]{Functionality Not Working \\ as Expected} & N/A & N/A & N/A & N/A & N/A  & \textbf{14} & N/A & \textbf{12} \\
  \midrule
  Wrong Gradient & N/A & N/A & N/A & N/A & N/A  & \textbf{1} & N/A & \textbf{1} \\
  \midrule
  Wrong Outputs & N/A & N/A & N/A & N/A & N/A  & \textbf{4} & N/A & \textbf{3} \\
  \bottomrule
  \end{tabular}
  \begin{tablenotes}
  \small
  \item[*] ``N/A'' indicates that the fuzzer is not applicable for detecting this specific type of bug.
  \end{tablenotes}
  \end{threeparttable}
  }
  \vspace{-10pt}
\end{table}

\subsection{RQ3: Ablation Study on Self-Validation}
\label{sec:verification}

To systematically assess the effectiveness of our \textit{LLM-Powered Self-Validation} in identifying real bugs, we conducted experiments using both real-world and curated datasets.
Our experiments focused on PyTorch as a representative case.

\noindent\textbf{False Positive Rate Analysis.}
In the real-world experiment presented in \S~\ref{sec:bugDiscovery}, \tool identified 31 previously unknown silent bugs. The detection process yielded a false positive rate of 28.58\%.
After further analysis, we identified the main reasons for \tool's misjudgments as follows:  
(1) \textit{Oracle design errors}: Logical flaws in the oracle cause correct behaviors to be misclassified as bugs.  
(2) \textit{Code migration failures}: The bug characteristics from the original code were not preserved or correctly mapped, leading to failures in reproducing the bug in other APIs.  
(3) \textit{Lack of information resulting in LLM misjudgment}: Due to insufficient information about the API, \tool fails to develop a deep understanding of its functionality.

While the absolute precision figure of 71.42\% may appear moderate in isolation, it is important to contextualize it within the extreme difficulty of automatically identifying and verifying silent bugs, a task for which general-purpose oracles are non-existent.
\tool is thus among the \textit{first} to systematically tackle this specific challenge using LLMs in DL library fuzzing.
The practical significance of this 71.42\% precision lies in its ability to serve as an effective filter.

\begin{figure}[h]
\centering
\includegraphics[width=\linewidth]{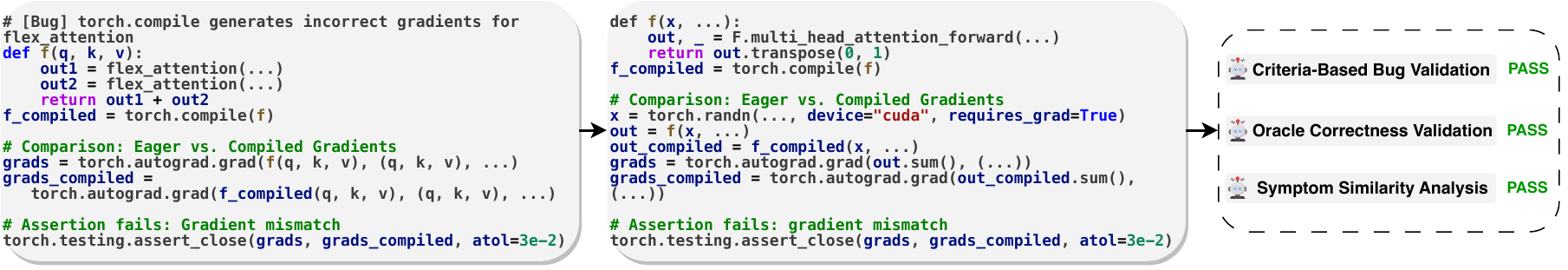}
\caption{Failure case passing all automated validations.}
\label{fig:example4}
\end{figure}

As shown in \F~\ref{fig:example4}, we present a failure case that passes all validations but is ultimately deemed non-bug after in-depth inspection by developers.
The original issue reports incorrect gradients produced by \textit{torch.compile} when a function composes multiple \textit{flex\_attention} calls, where the compiled gradients deviate from eager execution. Based on this pattern, we transferred the test to \textit{multi\_head\_attention\_forward}, again observing consistent gradient mismatches between eager and compiled executions.
The transferred case passed all automated validation stages and was initially labeled as \textit{high priority} by developers due to the apparent gradient inconsistency.
However, after several rounds of in-depth discussion and inspection, the developers concluded that the discrepancy arises from acceptable numerical differences introduced by compiler-level optimizations rather than a functional bug. These cases demand expert-level reasoning about compiler internals and floating-point behavior, exceeding the scope of our automated analysis.

\noindent \textbf{Comparison with Pure LLMs.}
To better understand the contribution of our system design, we conducted an ablation study on a curated dataset (detailed in Appendix~\ref{sec:ablationVer}). The results show that \tool's structured multi-step validation (achieving 0.8442 accuracy and a 0.7600 F1-score) is markedly superior to simpler LLM-only approaches. This underscores the importance of our system's design, rather than just the model's inherent capabilities. While the real-world complexity and incomplete API documentation pose limitations, the results indicate that \tool's design is key to its practical value in detecting elusive silent bugs.

\noindent\textbf{Different LLMs.}
To evaluate LLM performance in \textit{LLM-Powered Self-Validation}, we conducted comparative experiments on the curated dataset using leading mainstream models. These include the cost-effective model like GPT-4o mini~\cite{gpt4omini}, SOTA open-source models (DeepSeek R1~\cite{deepseekr1}, DeepSeek V3~\cite{deepseekv3}), and other commercial models (Doubao-1.5-thinking-pro~\cite{doubao}, Qwen-plus~\cite{qwen}). \T~\ref{tab:llmComparison} summarizes the results of different models on key metrics including accuracy, precision, recall, and F1 score.

As shown in the \T~\ref{tab:llmComparison}, model performance varies considerably. Both Doubao-1.5-thinking-pro and GPT-4.1 mini achieve high accuracy and recall, with F1 scores of 0.7556 and 0.76, respectively, demonstrating strong overall reasoning capabilities suitable for demanding scenarios. DeepSeek R1 leads in precision (0.7692) but shows lower recall. DeepSeek V3 and Qwen-plus display relatively balanced precision and recall. While GPT-4o mini is cost-effective, its low recall (0.2381) and F1 score hinder its suitability for this self-validation task.  
Strictness varies by model. GPT-4.1 mini adopts more lenient self-validation, achieving high recall (90.48\%) but lower precision (65.52\%), whereas more conservative models such as DeepSeek R1 enforce stricter validation, resulting in higher precision (76.92\%) but substantially lower recall (47.62\%).
This contrast shows that self-validation strictness is directly associated with the recall-precision trade-off: stricter validation leads to higher precision but lower recall, while more lenient validation yields higher recall at the cost of precision.

\begin{table}[h]
  \centering
  \caption{Performance comparison of different LLMs on the LLM-Powered Self-Validation task.}
  \vspace{-5pt}
  \label{tab:llmComparison}
  \resizebox{0.65\linewidth}{!}{
  \begin{tabular}{lcccc}
  \toprule
  \textbf{Model} & \textbf{Accuracy} & \textbf{Precision} & \textbf{Recall} & \textbf{F1 Score} \\
  \midrule
  GPT-4.1 mini             & 0.8442 & 0.6552 & 0.9048 & 0.7600 \\
  GPT-4o mini              & 0.7403 & 0.5556 & 0.2381 & 0.3333 \\
  DeepSeek R1              & 0.8182 & 0.7692 & 0.4762 & 0.5882 \\
  DeepSeek V3              & 0.7922 & 0.6087 & 0.6667 & 0.6364 \\
  Doubao-1.5-thinking-pro  & 0.8571 & 0.7083 & 0.8095 & 0.7556 \\
  Qwen-plus                & 0.7922 & 0.6316 & 0.5714 & 0.6000 \\
  \bottomrule
  \end{tabular}
  }
  \vspace{-5pt}
  \end{table}

\begin{figure}[t]
  \centering
  \begin{subfigure}[b]{0.495\textwidth}
      \centering
      \includegraphics[width=\textwidth]{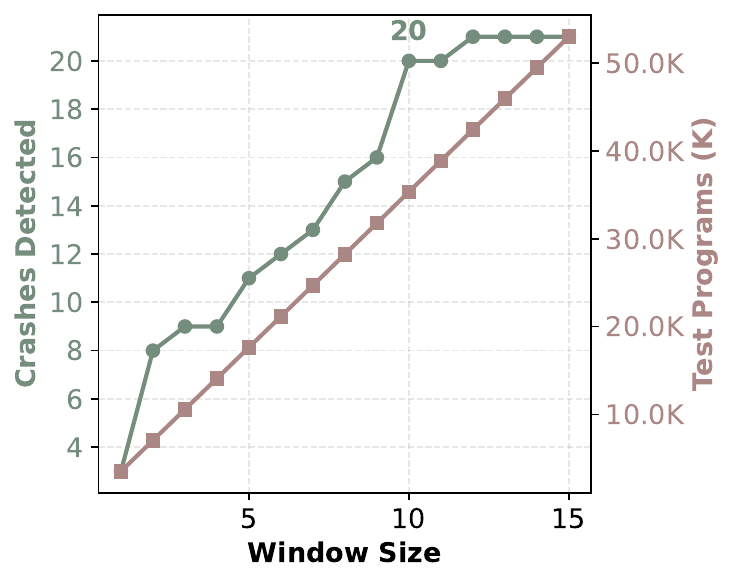}
      \vspace{-15pt}
      \caption{Effect of window size on bug discovery and test program count in
      PyTorch$\rightarrow$PyTorch bug transfer.}
      \vspace{-10pt}
      \label{fig:window}
  \end{subfigure}
  \hfill
  \begin{subfigure}[b]{0.495\textwidth}
      \centering
      \includegraphics[width=\textwidth]{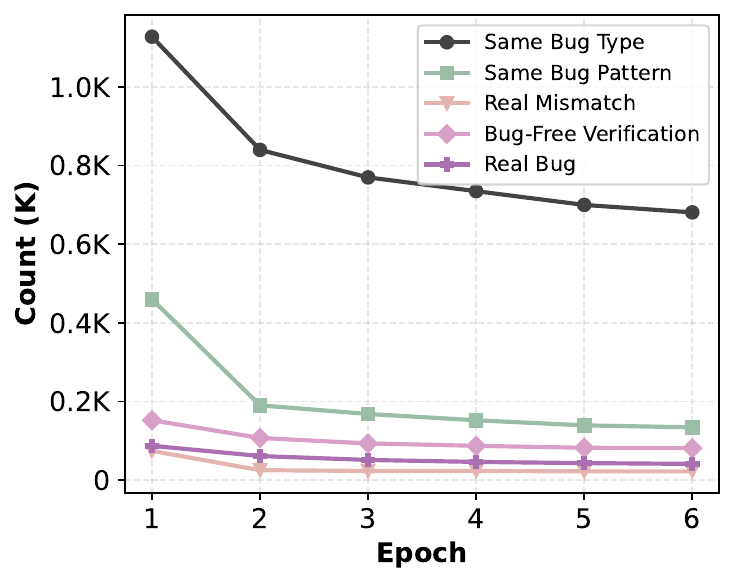}
      \vspace{-15pt}
      \caption{Number of cases determined as ``pass'' under different numbers of repeated checks per prompt.}
      \vspace{-10pt}
      \label{fig:repeat}
  \end{subfigure}
  
  \caption{Ablation study on hyperparameters.}
  \label{fig:hyper}
  \vspace{-10pt}
  \end{figure}

\subsection{RQ4: Ablation Study on Hyperparameters}
\label{sec:hyperparameter}
\noindent\textbf{Window Size.} 
In \textit{Bug Transfer-Guided Test Generation}, the exploration phase adopts a batch testing strategy, selecting 10 untested APIs per round most similar to the original bug API based on the similar API queue. If any new bugs are found in a batch, testing proceeds to the next round; otherwise, it terminates early.
To evaluate the impact of window size (i.e., the number of APIs tested per round) on effectiveness and resource usage, we conducted a hyperparameter sensitivity study using the ``PyTorch$\rightarrow$PyTorch'' transfer scenario.
For each original buggy API, we attempted transfers to the top 30 most similar target APIs, varying the window size from 1 to 15.
For each setting, we recorded both the number of generated test programs and the number of crashes found.

As shown in \F~\ref{fig:window}, increasing the window size initially yields more crashes, but the benefit plateaus beyond a size of 10. In contrast, test case volume grows linearly with window size. At a window size of 10, 20 out of 21 total crashes are already discovered, while test case count (35,350) remains significantly lower than that at size 15 (53,010).
These results suggest that a window size of 10 offers a strong balance between effectiveness and cost, enabling high bug discovery with reduced redundant testing.


\noindent\textbf{Repeated Time.}
In the \textit{LLM-Powered Self-Validation} process, LLMs are used to automatically analyze test programs and distinguish real bugs from false positives. However, due to the inherent randomness in LLM outputs, repeated prompts can yield inconsistent results, potentially undermining validation stability.
To reduce misjudgments caused by randomness, we repeat each check prompt multiple times. The verdict rule for \textit{LLM-Powered Self-Validation} is: as long as any repetition fails, the case is considered to have failed (i.e., result = epch1\_res \& epch2\_res \& epch3\_res...).
Our goal is to determine how many repetitions should be used for each prompt, so as to minimize misjudgment due to random fluctuations, while also managing the token consumption.

We selected 5 prompts from \textit{LLM-Powered Self-Validation} to filter out false positives for our experiments.
For each prompt, we applied 1 to 6 repeats and count the number of cases
finally ``judged as passed''. \F~\ref{fig:repeat} shows the results. In
the initial stage (1--3 repeats), the descent is the steepest; after 3 repeats,
the results enter a stable phase and marginal changes diminish. For example, in
prompt \textit{Real Mismatch?}, there is almost no change after the third repetition; in prompt \textit{Same Bug Type?}, the
decrease after the third repetition is much smaller; other prompts show the same
trend. Therefore, we finally set the repeat number to 3 to maximize the
balance between reducing randomness and controlling token consumption.



\subsection{RQ5: Ablation Study on Prompt Types}
\label{sec:promptAnalysis}
We conducted a manual analysis of the key prompts to evaluate the LLM's performance across various sub-tasks. 
We randomly selected 10 instances for each prompt. Note that the original issues corresponding to these cases passed the filtering checks of criteria-based bug validation, ensuring that the analyzed issues are of high value.

As shown in Table~\ref{tab:manual}, six tasks, including \textit{Extraction of Bug Determination Criteria}, achieve over 70\% accuracy, indicating that LLMs are well-suited for these tasks. 
However, for the task \textit{Real Bug?}, the accuracy drops to just 40\%. 
We attribute this performance gap to the increased complexity of these tasks, which require LLMs to perform intricate synthesis and reasoning across multiple types of information, making them inherently challenging. To tackle this problem, we adopt a debate mechanism where the LLM questions its initial results from an opposing perspective, ultimately summarizing the findings, which improves precision to 71.42\% (see \S~\ref{sec:verify}).
The relatively low accuracy in \textit{Context-Aware Bug Transfer} is due to the fact that the target APIs identified through functionality-based API matching may include APIs with substantial differences. 
To address this issue, we dynamically adjust the target APIs based on actual fuzzing results (see \S~\ref{sec:transferFuzzing}).

\begin{table}[h]
\centering
\caption{Performance of LLM on different prompt types.}
\vspace{-5pt}
\resizebox{0.6\linewidth}{!}{
\begin{tabular}{l|c}
\toprule
\textbf{Prompt Type}         & \textbf{Accuracy} \\
\midrule

Context-Aware Bug Pattern (\F~\ref{fig:issueAnalysis})    & 90\%              \\
Context-Free Functionality Description (\F~\ref{fig:apiAnalysis}) & 100\%              \\
Context-Aware Bug Transfer (\F~\ref{fig:bugTransfer}) & 60\%              \\
Same Bug Type? (\F~\ref{fig:sameBugType})                & 90\%              \\
Real Mismatch? (\F~\ref{fig:realMismatch})                & 100\%             \\
Same Bug Pattern? (\F~\ref{fig:sameBugPattern})             & 70\%              \\
Bug-free Verification (\F~\ref{fig:bugFree})        & 40\%              \\
Extraction of Bug Determination Criteria (\F~\ref{fig:insightAnalysis})      & 100\%             \\
Real Bug? (\F~\ref{fig:insightCheck})             & 40\%              \\

\bottomrule
\end{tabular}
}
\label{tab:manual}
\vspace{-10pt}
\end{table}

\subsection{RQ6: Cost Analysis}
To evaluate the efficiency and cost-effectiveness of \tool, we report the model selection and the cost associated with each key component in our pipeline. \T~\ref{tab:modelCost} summarizes the models used and the total expenditure for each submodule throughout the entire evaluation in the ``PyTorch$\rightarrow$PyTorch'' setting.

The \textit{Context-Aware Bug Pattern Extraction} phase accounts for the largest portion of the total cost (\$42.16; 47.33\%), which is expected given the complexity involved in analyzing a large number of issue reports. This high expenditure is justified by the need for accuracy when extracting context-aware bug patterns from diverse and often lengthy issue descriptions.
\textit{Bug Transfer-Driven Fuzzing} is the second most expensive component (\$27.60; 30.98\%), as it involves generating 35,909 test programs and designing context-aware trigger conditions and oracles for each one.
The \textit{LLM-Powered Self-Validation} step, while relatively less costly (\$18.99; 21.32\%), is crucial for ensuring the correctness of transferred bugs and identifying potential false positives, thereby enhancing the overall reliability of our system.
In contrast, \textit{Functionality-Based API Matching} incurs a negligible cost (\$0.32; 0.36\%) due to the simplicity of the task.
Overall, the total monetary cost for running \tool on PyTorch is \$89.07.

\begin{table}[t]
\centering
\caption{Model selection and cost breakdown for each component.}
\vspace{-5pt}
\resizebox{0.75\linewidth}{!}{

\begin{tabular}{l|l|r|r}
\toprule
\textbf{Component} & \textbf{Model Used} & \textbf{Cost (USD)} & \textbf{Proportion (\%)} \\
\midrule
Context-Aware Bug Pattern Extraction  & o3-mini      & \$42.16 & 47.33 \\
Functionality-Based API Matching  & GPT-4o mini  & \$0.32  & 0.003 \\
Bug Transfer-Driven Fuzzing & GPT-4o mini  & \$27.60 & 30.98 \\
LLM-Powered Self-Validation & GPT-4.1 mini & \$18.99 & 21.32 \\
\midrule
\textbf{Total} & --           & \textbf{\$89.07} & \textbf{100} \\
\bottomrule
\end{tabular}

}
\vspace{-10pt}

\label{tab:modelCost}
\end{table}

\section{Discussion}



\tool leverages triggering contexts and oracles derived from historical bug reports to guide the detection of similar silent bugs. Because silent bugs are highly context-dependent, their detection often requires manual, case-by-case oracle design, making them largely undetectable by existing methods.
As the first work to automate silent bug testing in deep learning libraries, our primary goal is to build a practical and effective testing framework rather than exhaustively eliminating all false negatives, which is generally impossible in software testing.

\tool is designed to automatically detect silent bugs while minimizing false positives, a critical bottleneck in existing DL library testing tools. This emphasis inevitably introduces the risk of filtering out some real bugs (i.e., false negatives). To balance this trade-off, we adopt a ``transfer-then-verify'' strategy. In the transfer stage, we aim for broad coverage to surface as many potential anomalies as possible. In the subsequent validation stage, we apply conservative filtering to remove invalid reports while retaining suspicious cases.
Our manual inspection of the misclassified samples in \T~\ref{tab:manual} shows that the observed errors correspond to misclassifying expected behavior as real bugs (false positives), with no cases where confirmed real bugs were filtered out. While this does not imply that false negatives are fully eliminated, it suggests that the validation is conservative in practice and does not aggressively discard potential bugs. Further optimizations will be explored in future work.

\section{Conclusion}
In this paper, we presented \tool, a transfer-then-verify framework designed for silent bug fuzzing in deep learning libraries.
By leveraging LLMs to extract context-aware bug patterns, match semantically related APIs using functionality-based embeddings, and synthesize tests with customized oracles, \tool enables versatile yet controlled bug transfer. To ensure precision, it integrates an LLM-powered self-validation module to reduce false positives.
Evaluations show that \tool uncovers 79 previously unknown bugs across PyTorch, TensorFlow, and MindSpore, demonstrating its effectiveness and cross-framework generalizability.

\section*{Data-Availability Statement}
The code and datasets used in this work are available at \url{https://github.com/transfuzz/transfuzz}. While we are committed to sharing these materials openly, we will not be submitting the artifact for formal Artifact Evaluation.

\section*{Acknowledgments}
The HKUST authors were supported in part by a RGC GRF grant under the contract 16214723, an ITF grant under the contract ITS/161/24FP, a HKUST Bridge The Gap fund BGF.001.2025.

\bibliographystyle{ACM-Reference-Format}
\bibliography{bib/ref,bib/similarity,bib/decompiler,bib/machine-learning,bib/attack}

\appendix

\begin{table}[htbp]
  \centering
  \vspace{-5pt}
  \caption{Discovered CVEs in three DL libraries and their CVSS 4.0 base scores. To comply with the double-blind policy, specific CVE numbers are omitted.}
  \label{tab:cve}
  \resizebox{0.55\linewidth}{!}{
  \begin{tabular}{ll}
  \toprule
  \multicolumn{2}{c}{CVE ID} \\
  \midrule
  CVE-2025-2148 (2.3 LOW) & CVE-2025-2998 (4.8 MEDIUM) \\
  CVE-2025-2149 (2.0 LOW) & CVE-2025-2999 (4.8 MEDIUM) \\
  CVE-2025-2953 (4.8 MEDIUM) & CVE-2025-3000 (4.8 MEDIUM) \\
  CVE-2025-3001 (4.8 MEDIUM) & CVE-2025-3121 (4.8 MEDIUM) \\
  CVE-2025-3136 (4.8 MEDIUM) & CVE-2025-3144 (4.8 MEDIUM) \\
  CVE-2025-3730 (4.8 MEDIUM) & CVE-2025-4287 (4.8 MEDIUM) \\
  \bottomrule
  \end{tabular}
  }
  \end{table}

\section{Oracles Distribution}
\label{sec:oracleCount}
\begin{figure}[h]
  \centering
  \includegraphics[width=\linewidth]{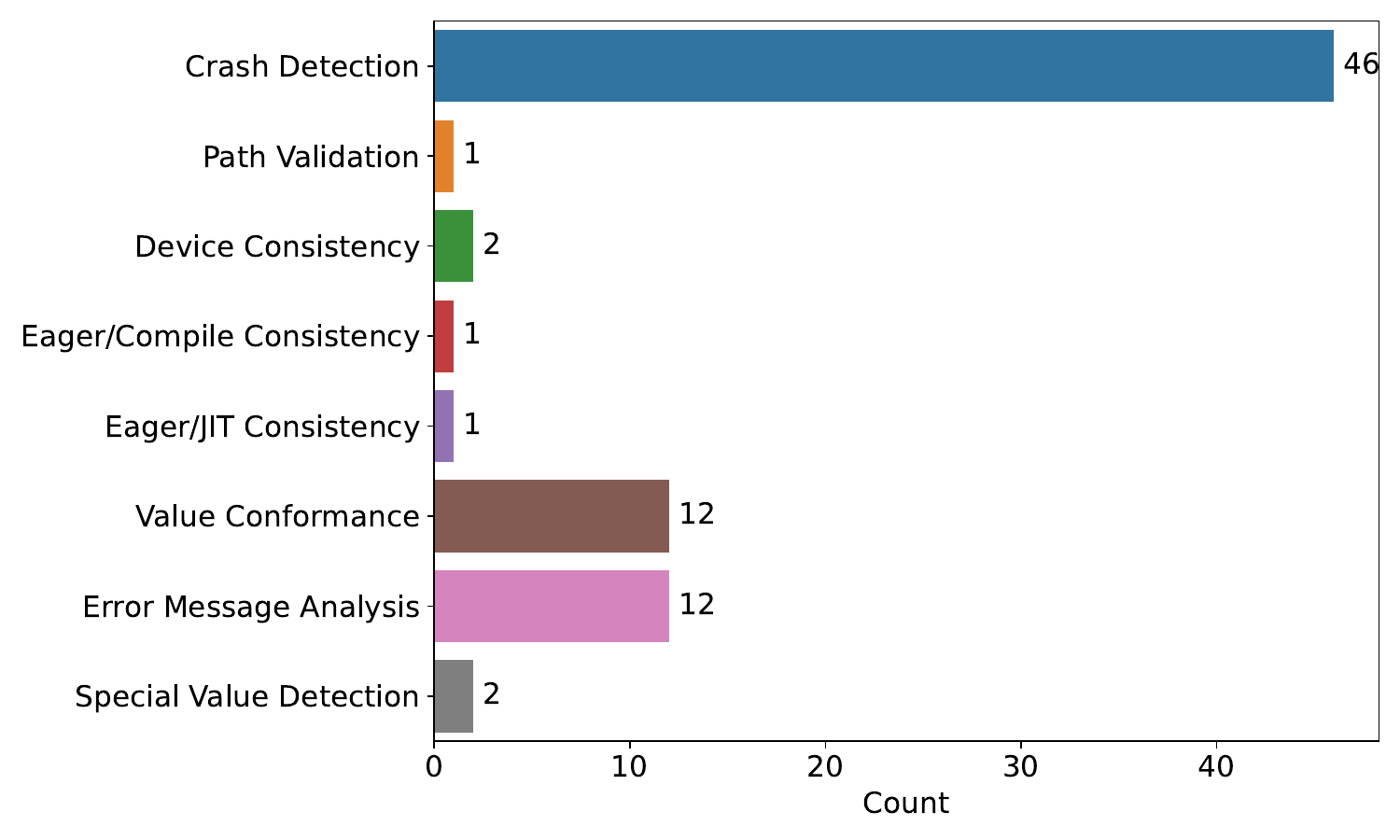}
  \caption{Distribution of oracle types used to detect the bugs listed in \T~\ref{tab:silent} and \ref{tab:crash}, showing the frequency of each oracle type.}
  \label{fig:oracleCount}
\end{figure}

We also analyze the distribution of oracle types used by \tool to detect the bugs listed in \T~\ref{tab:silent} and \T~\ref{tab:crash}, and the results are shown in \F~\ref{fig:oracleCount}.
Among them, \textit{Crash Detection} accounts for the largest portion (46 cases), indicating that program crashes remain the easiest type of bug to trigger and are the most representative.  
\textit{Value Conformance} (12 cases) highlights the core challenge faced by DL libraries in ensuring numerical computation reliability.  
\textit{Error Message Analysis} (12 cases) and \textit{Special Value Detection} (2 cases) capture non-crash bugs such as misleading warnings or unexpected special values.  
\textit{Device Consistency} (2 cases), \textit{Eager/Compile Consistency} (1 case), and \textit{Eager/JIT Consistency} (1 case) reflect the need to verify consistency across hardware and execution environments, not only in output values but also in performance behavior.  

As an example of \textit{Device Consistency}, \tool can detect performance degradation across devices.
Given a known bug where \texttt{torch.linalg.svd} exhibits a significant GPU performance regression compared to CPU, \tool transfers this bug pattern to \texttt{torch.functional.pca\_lowrank}, which implements a similar matrix factorization algorithm, and constructs an oracle that compares GPU and CPU execution time, reporting a bug when GPU execution is an order of magnitude slower.
These results highlight that \tool can automatically design diverse oracles based on bug behavior to detect different types of bugs.

\section{Semantic Transfer Patterns for Silent Bug Discovery}
\label{sec:apiRelationship}
In this section, we present an empirical analysis of semantic-level bug transfer enabled by \tool. \T~\ref{tab:semantic-transfer-types} summarizes representative silent bugs discovered via \tool through semantic-level transfer across APIs. In each case, the source and target APIs do not merely share naming or version-level similarity; instead, they exhibit a deeper semantic commonality, such as functional similarity, shared mathematical domains, type inference logic, or implementation-level dependencies.
Notably, most transfers are non-trivial: the involved APIs often reside in different modules or serve distinct usage scenarios, yet expose similar latent bugs due to shared semantics. This demonstrates that \tool goes beyond testing ``parallel APIs'' and instead leverages LLM-guided reasoning to identify and exploit deep logical relationships between APIs, enabling effective bug transfer across diverse API boundaries.

\begin{table}[h]
\centering
\caption{Nine semantic transfer types identified by TransFuzz. Each type is defined by a shared semantic basis between the source and target APIs, enabling systematic bug transfer beyond API name similarity.}
\label{tab:semantic-transfer-types}
\small
\resizebox{\linewidth}{!}{
\begin{tabular}{lll}
\toprule
\textbf{Transfer Type} &
\textbf{Semantic Basis} &
\textbf{Example API Pair} \\
\midrule
Functional Similarity
& Similar core operation (log-softmax)
& \texttt{log\_softmax $\rightarrow$ sparse.log\_softmax} \\

Mathematical Domain
& Similar mathematical domain (inverse hyperbolic)
& \texttt{acosh $\rightarrow$ asinh\_} \\

Type System Semantics
& Similar type inference logic (Python$\rightarrow$Tensor)
& \texttt{tensor $\rightarrow$ default\_collate} \\

Implementation Kernel
& Similar algorithmic kernel (matrix factorization)
& \texttt{svd $\rightarrow$ pca\_lowrank} \\

Runtime System
& Similar warning generation mechanism
& \texttt{compile $\rightarrow$ jit.trace} \\

Design Pattern
& Similar dynamic class construction pattern
& \texttt{OpaqueUnaryFn $\rightarrow$ BitwiseFn} \\

Usage State
& Similar persistent quantized state
& \texttt{quant.Hardswish $\rightarrow$ quant.Sigmoid} \\

Input Semantics
& Similar NumPy-to-Tensor conversion path
& \texttt{tensor $\rightarrow$ default\_collate} \\

Argument Contract
& Similar argument semantics (\texttt{out=})
& \texttt{matmul $\rightarrow$ tensordot} \\
\bottomrule
\end{tabular}
}
\end{table}

\section{Ablation Study on the Components of LLM-Powered Self-Validation}
\label{sec:ablationVer}
To investigate whether the performance gains of TransFuzz stem from the system design of \textit{LLM-Powered Self-Validation} rather than the general capabilities of LLMs, we construct an additional evaluation dataset. The dataset is derived from an early TransFuzz prototype without self-validation, where alarm-triggering cases were directly judged by an LLM, resulting in low precision due to the absence of explicit bug determination criteria. We manually curated 78 representative and complex cases (22 real bugs, 56 non-bugs) from this process.

Importantly, this dataset is used only to compare the relative effectiveness of different validation strategies, rather than to demonstrate absolute performance. While \tool achieves high accuracy on this curated dataset, it also maintains high precision in real-world testing on the latest PyTorch version (See \S~\ref{sec:verification}). This consistency suggests that the benefits of our self-validation design generalize beyond the constructed evaluation set.

Our \textit{LLM-Powered Self-Validation} system consists of two principal components: (1) Instrumentation and (2) Bug transfer validation. 
We focus our comparison on the following methods:
\begin{itemize}
    \item LLM-only: The LLM directly analyzes the test program source code and execution outputs to determine whether a real bug is triggered.
    \item Instrumentation + LLM: Based on the results of program instrumentation, the LLM is provided with both source code and detailed execution information for bug determination.
    \item \tool: Leverages both instrumentation and the full bug transfer validation.
\end{itemize}

We evaluate the three methods using four metrics: Accuracy, Precision, Recall, and F1 Score. The definitions of these metrics are summarized as follows:

\[
\begin{aligned}
&\text{Accuracy} = \frac{TP + TN}{TP + FP + TN + FN} \\
&\text{Precision} = \frac{TP}{TP + FP} \\
&\text{Recall} = \frac{TP}{TP + FN} \\
&\text{F1~Score} = \frac{2 \cdot \text{Precision} \cdot \text{Recall}}{\text{Precision} + \text{Recall}}
\end{aligned}
\]

\noindent where \( TP \), \( FP \), \( TN \), and \( FN \) denote the numbers of true positives, false positives, true negatives, and false negatives, respectively.

The quantitative results on the constructed dataset are summarized in \T~\ref{tab:datasetRes}. The results reveal that the naive LLM-only approach, when directly applied, suffers from a high false positive rate, leading to low precision (0.3077) despite its decent recall (0.7619). Incorporating instrumentation mitigates the false positive rate and improves specificity, but at the cost of reduced recall. Our method (\tool), integrating runtime feedback with multi-step bug transfer validation, substantially outperforms the baselines across all major metrics: it achieves the highest accuracy (0.8442), precision (0.6552), recall (0.9048), and F1 score (0.7600). 
Notably, our approach maintains high recall while greatly increasing precision, demonstrating its effectiveness in suppressing false positives and reliably identifying real bugs. This highlights the benefit of our system design beyond the raw capabilities of LLMs.

\begin{table}[h]
\caption{Quantitative results over the constructed dataset.}
\label{tab:datasetRes}
\centering
\begin{tabular}{lcccc}
\toprule
Method                  & Accuracy & Precision & Recall  & F1 Score \\
\midrule
LLM-only                & 0.4675   & 0.3077    & 0.7619  & 0.4384   \\
Instrumentation + LLM   & 0.6494   & 0.3846    & 0.4762  & 0.4255   \\
\tool      & 0.8442   & 0.6552    & 0.9048  & 0.7600   \\
\bottomrule
\end{tabular}
\end{table}

\section{IR Definitions for Bug Types}
\label{sec:bugIR}
In \S~\ref{sec:verify}, we use IR to rigorously define common bug types, thereby transforming the subjective problem of whether bugs are similar into a classification problem of bug types. Here, we provide the IR definitions for the remaining six common bug types:
{\footnotesize
\[
\begin{aligned}
\text{JIT\_Eager\_Mismatch} ::= \\
&\hspace{-8em} \texttt{APICall(api)[mode=eager]} \rightarrow v_1 \wedge \\
&\hspace{-8em} \texttt{APICall(api)[mode=(jit\_trace|jit\_script)]} \rightarrow v_2 \wedge \\
&\hspace{-8em} \texttt{OracleCheck(ValueCorrectness)}( \\
&\hspace{-6em} \texttt{condition=Compare}(v_1, v_2) \\
&\hspace{-8em}) \rightarrow \texttt{FAIL}
\end{aligned}
\]

\[
\begin{aligned}
\hspace{-1em}\text{Compile\_Eager\_Mismatch} ::= \\
&\hspace{-10em} \texttt{APICall(api)[mode=eager]} \rightarrow v_1 \wedge \\
&\hspace{-10em} \texttt{APICall(api)[mode=compile(fx|dynamo)]} \rightarrow v_2 \wedge \\
&\hspace{-10em} \texttt{OracleCheck(ValueCorrectness)}( \\
&\hspace{-8em} \texttt{condition=Compare}(v_1, v_2) \\
&\hspace{-10em}) \rightarrow \texttt{FAIL}
\end{aligned}
\]

\[
\begin{aligned}
\hspace{-7em}\text{Functional\_Defect} ::= \\
&\hspace{-6.5em} \texttt{APICall(api)} \rightarrow v_5 \wedge \\
&\hspace{-6.5em} \texttt{OracleCheck(ValueCorrectness)}( \\
&\hspace{-4.5em} \texttt{condition=MatchValue}(v_5, \textit{expected}), \\
&\hspace{-4.5em} \texttt{criteria} = [\textit{dtype}, \textit{shape}, \textit{numerical}] \\
&\hspace{-6.5em}) \rightarrow \texttt{MISMATCH}
\end{aligned}
\]

\[
\begin{aligned}
\hspace{-8em}\text{Execution\_Crash} ::= \\
&\hspace{-7em} \texttt{APICall(api)[mode=*]} \rightarrow \textit{fault} \wedge \\
&\hspace{-7em} \texttt{OracleCheck(ExceptionType)}( \\
&\hspace{-5em} \texttt{condition=FaultType}(\textit{fault}) \in \{ \\
&\hspace{-5em} \textit{FloatingPointException}, \textit{SegFault}, \textit{Aborted} \} \\
&\hspace{-7em}) \rightarrow \texttt{TRIGGERED}
\end{aligned}
\]

\[
\begin{aligned}
\hspace{-8.1em}\text{Precision\_Degradation} ::= \\
&\hspace{-9.2em} \texttt{APICall(api)} \rightarrow v \wedge \\
&\hspace{-9.2em} \texttt{OracleCheck(ValueCorrectness)}( \\
&\hspace{-7.2em} \texttt{condition=Compare}(v, \textit{expected}), \\
&\hspace{-7.2em} \texttt{tolerance} = \{ \text{abs}:10^{-4}, \text{rel}:10^{-5} \} \\
&\hspace{-9.2em}) \rightarrow \texttt{FAIL}
\end{aligned}
\]

\[
\begin{aligned}
\text{Security\_Risk} ::= \\
&\hspace{-7em} (\texttt{VarDef} \mid \texttt{APICall} \mid \texttt{ControlBlock})[\text{security\_sensitive}=\text{true}] \wedge \\
&\hspace{-7em} \texttt{OracleCheck(SecurityViolation)}( \\
&\hspace{-5em} \texttt{condition=SecurityPolicyCheck}( \\
&\hspace{-5em}quad\ \textit{violation\_type} = (\textit{Code Injection Attack} \mid \ldots), \\
&\hspace{-5em}quad\ \textit{severity} = (\textit{high} \mid \textit{medium} \mid \textit{low}) \\
&\hspace{-5em} ), \\
&\hspace{-5em} \texttt{capture\_tensors} = [\textit{security\_relevant\_vars} \ldots] \\
&\hspace{-7em}) \rightarrow \texttt{VIOLATION\_DETECTED}
\end{aligned}
\]

}

\section{All Prompts used in \tool}
\label{sec:prompts}

\begin{figure}
  \centering
  \includegraphics[width=0.6\linewidth]{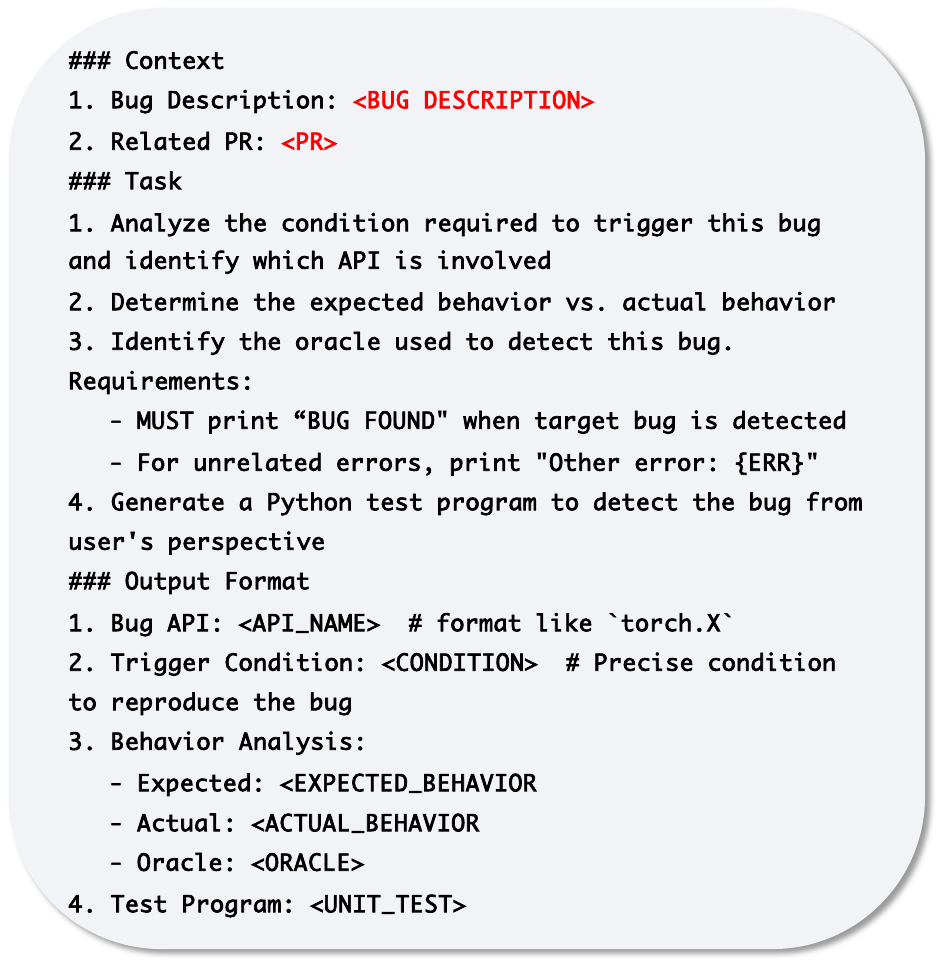}
  \caption{Prompt for context-aware bug pattern extraction.}
  \label{fig:issueAnalysis}
\end{figure}

\begin{figure}
  \centering
  \includegraphics[width=0.6\linewidth]{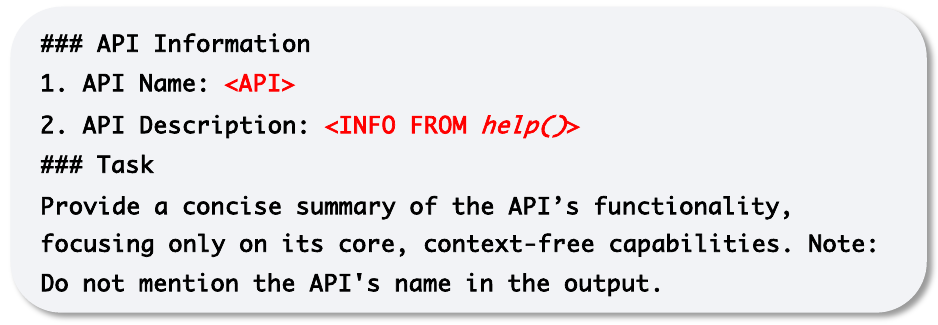}
  \caption{Prompt for the extraction of context-free functional descriptions.}
  \label{fig:apiAnalysis}
\end{figure}

\begin{figure}
  \centering
  \includegraphics[width=0.6\linewidth]{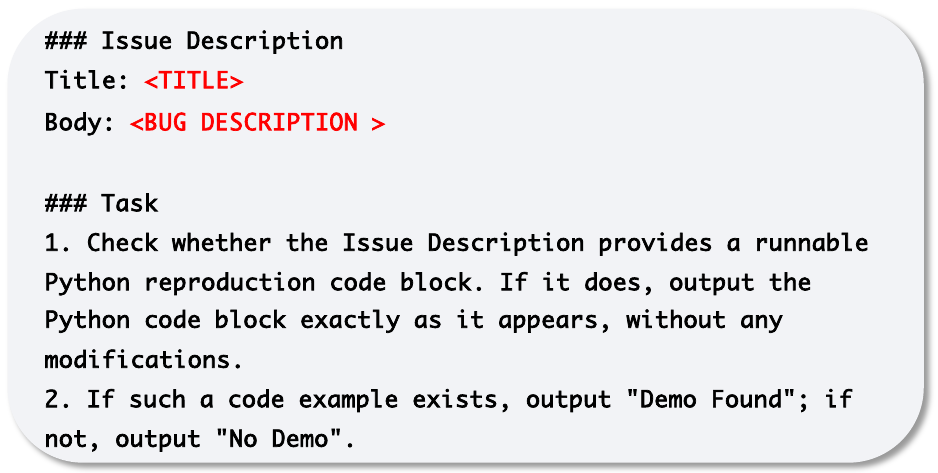}
  \caption{Prompt to determine whether an issue provides reproduction code.}
  \label{fig:issueCheck2}
\end{figure}

\begin{figure}
  \centering
  \includegraphics[width=0.6\linewidth]{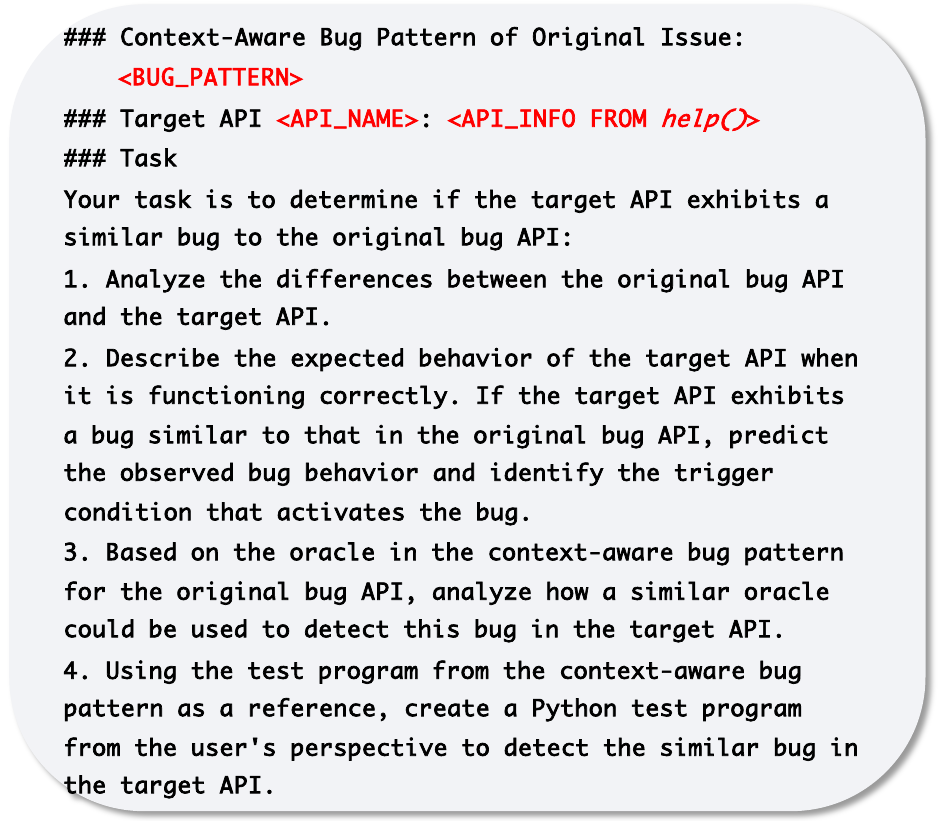}
  \caption{Prompt for context-aware bug transfer.}
  \label{fig:bugTransfer}
\end{figure}

\begin{figure}
  \centering
  \includegraphics[width=0.6\linewidth]{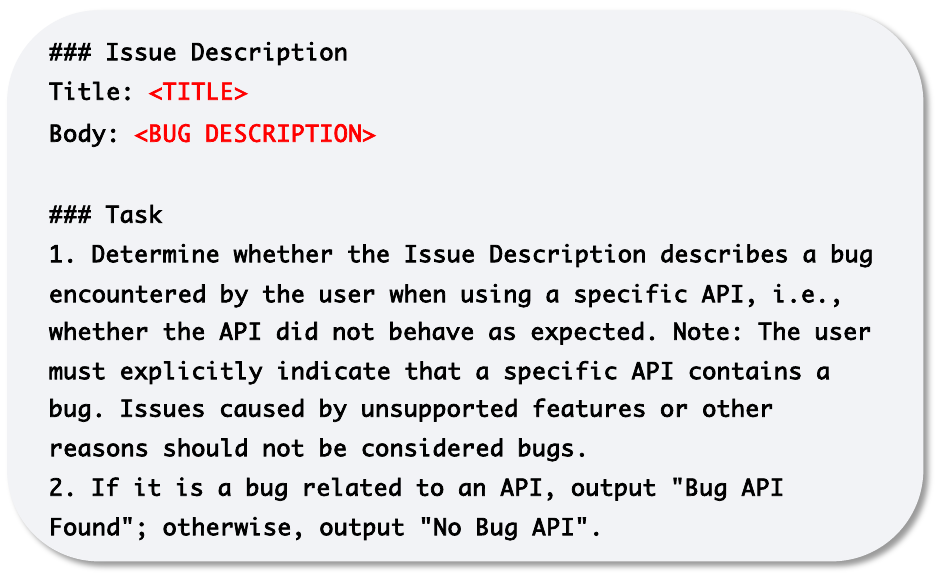}
  \caption{Prompt to determine if an issue describes an API bug.}
  \label{fig:issueCheck1}
\end{figure}

\begin{figure}
  \centering
  \includegraphics[width=0.6\linewidth]{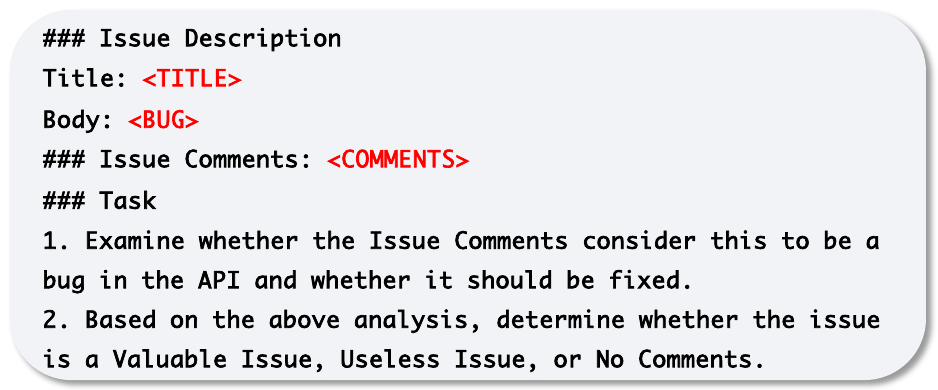}
  \caption{Prompt for analyzing issue comments.}
  \label{fig:issueCheck3}
\end{figure}

\begin{figure}
  \centering
  \includegraphics[width=0.6\linewidth]{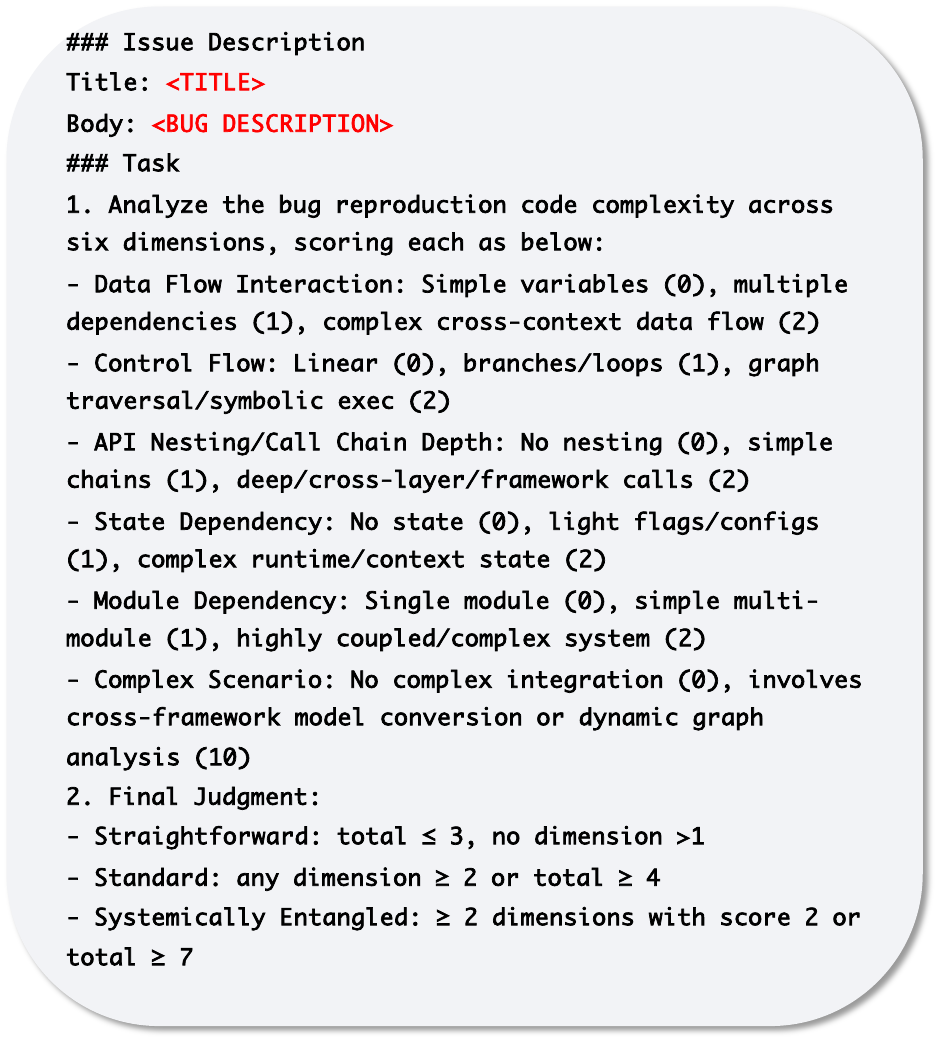}
  \caption{Prompt for analyzing the complexity of reproduction code.}
  \label{fig:issueCheck4}
\end{figure}

\begin{figure}
  \centering
  \includegraphics[width=0.6\linewidth]{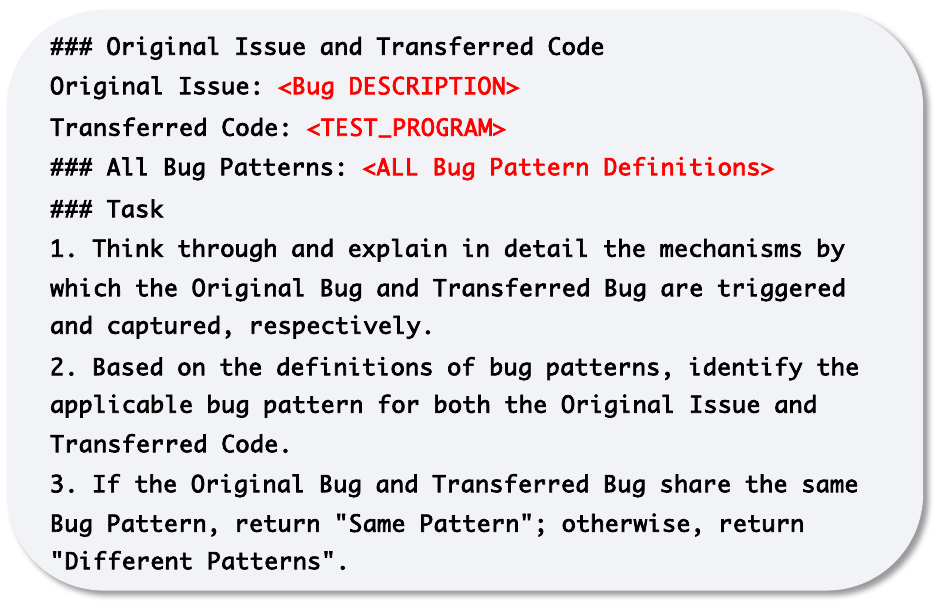}
  \caption{Prompt for answering \textit{Same Bug Type?}}
  \label{fig:sameBugType}
\end{figure}


\begin{figure}
  \centering
  \includegraphics[width=0.6\linewidth]{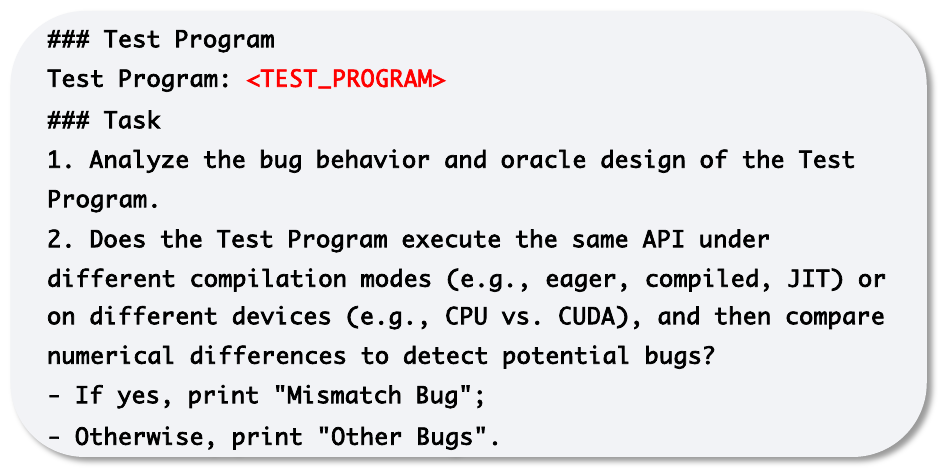}
  \caption{Prompt for answering \textit{Real Mismatch Bug?}}
  \label{fig:realMismatch}
\end{figure}

\begin{figure}
  \centering
  \includegraphics[width=0.6\linewidth]{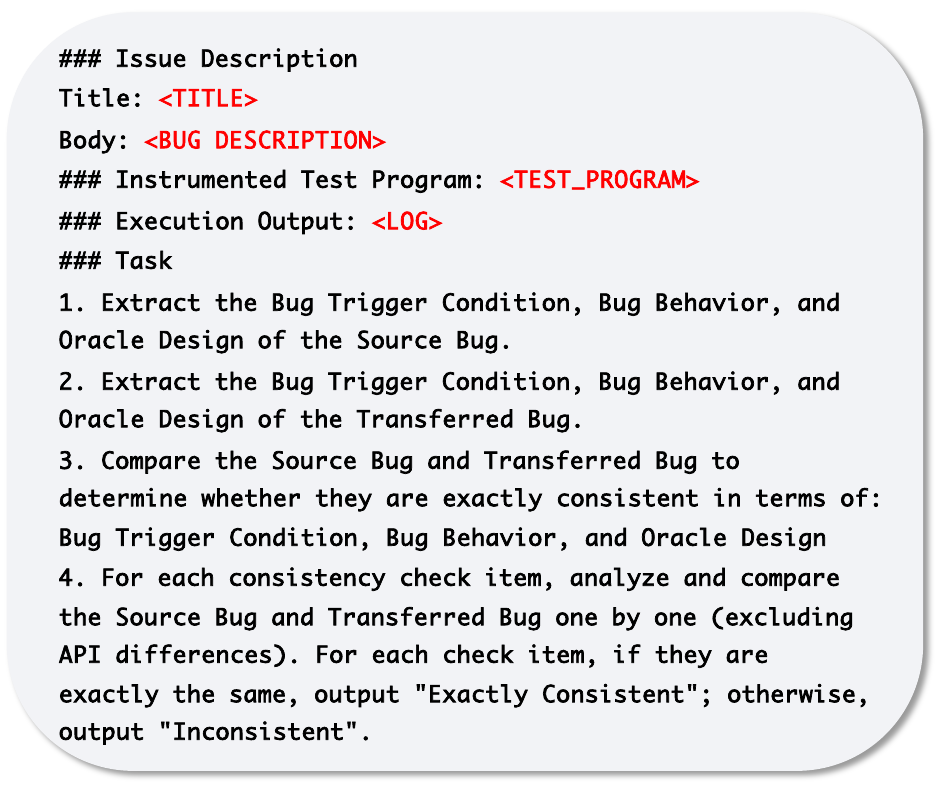}
  \caption{Prompt for answering the \textit{Same Bug Pattern?}}
  \label{fig:sameBugPattern}
\end{figure}

\begin{figure}
  \centering
  \includegraphics[width=0.6\linewidth]{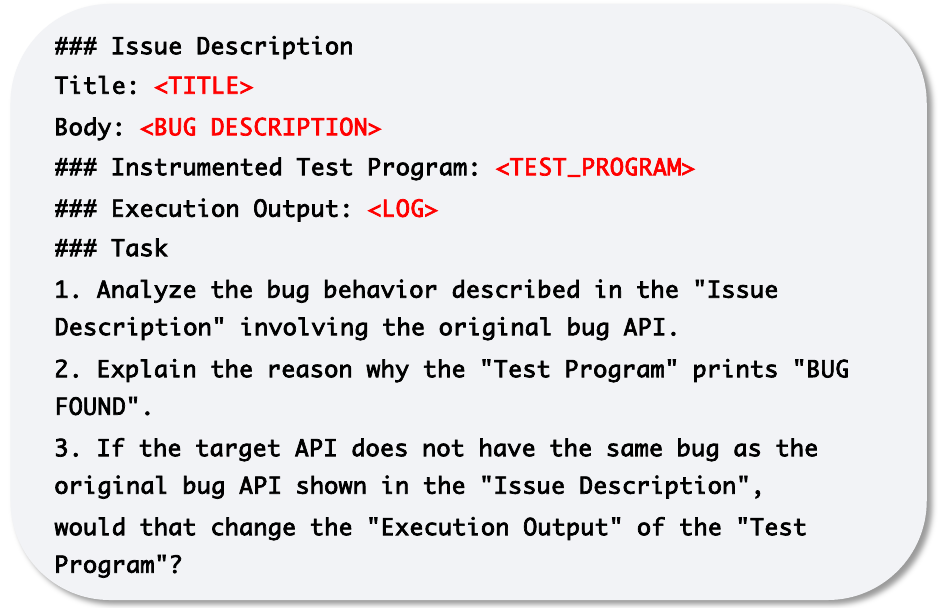}
  \caption{Prompt for the analysis of \textit{Bug-free Verification}.}
  \label{fig:bugFree}
\end{figure}

\begin{figure}
  \centering
  \includegraphics[width=0.6\linewidth]{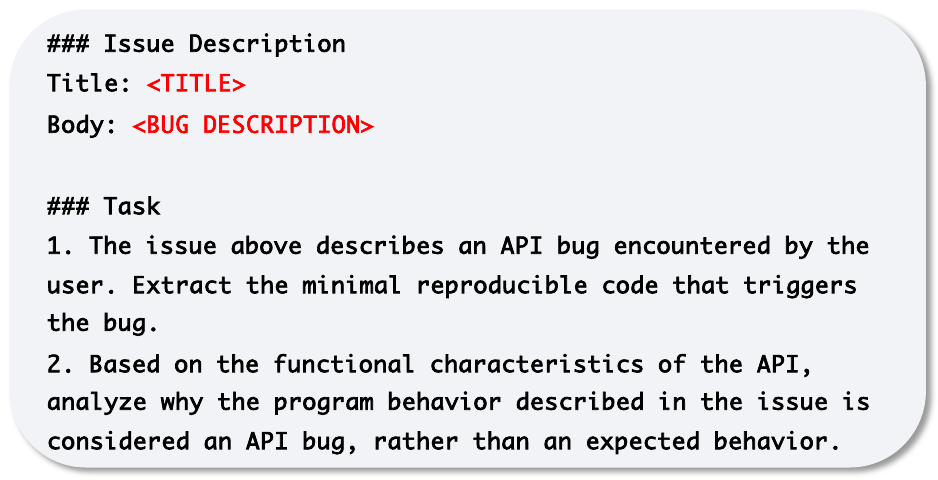}
  \caption{Prompt for the \textit{Extraction of Bug Determination Criteria}.}
  \label{fig:insightAnalysis}
\end{figure}

\begin{figure}
  \centering
  \includegraphics[width=0.6\linewidth]{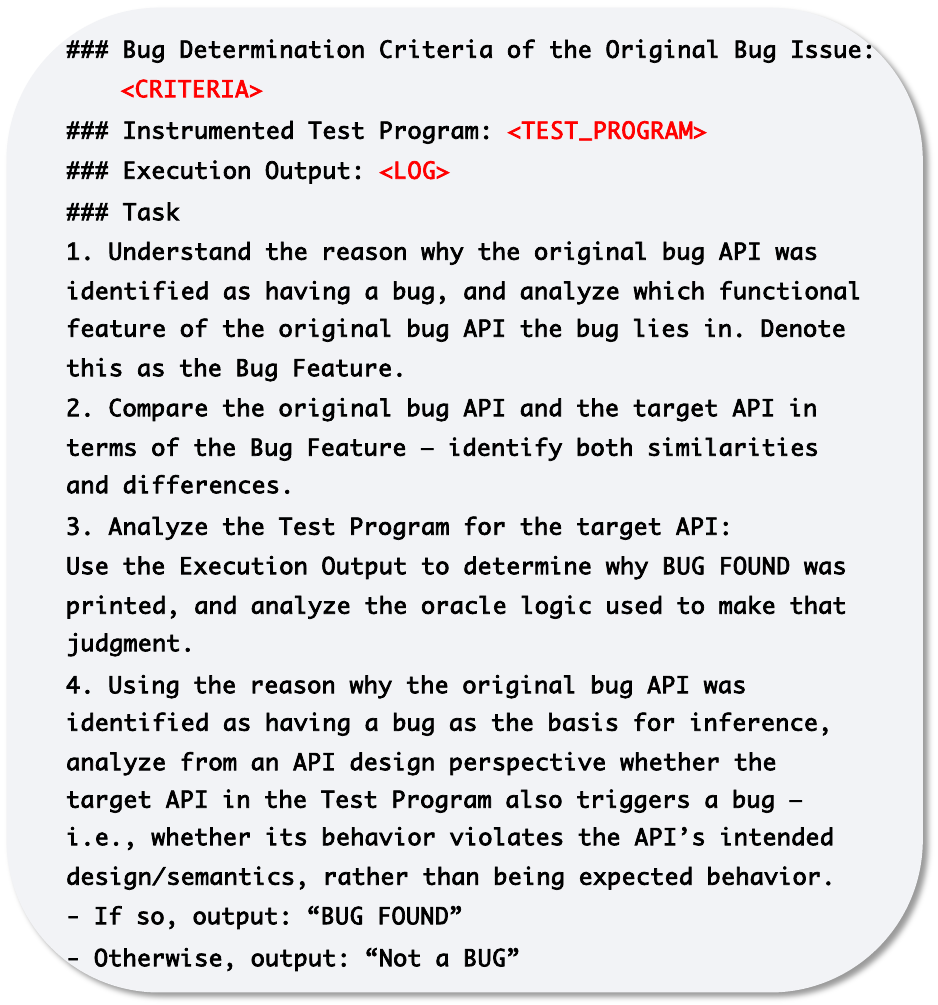}
  \caption{Prompt for answering \textit{Real Bug?}}
  \label{fig:insightCheck}
\end{figure}

\end{document}